\documentclass[preprint]{aastex61}
\usepackage{graphicx}
\usepackage{multirow}
\usepackage{amsmath}
\usepackage{amssymb}
\usepackage{gensymb}

\begin{document}
\title{Interactions Between Radio Galaxies and Cluster Shocks - 1: Jet Axes Aligned with Shock Normals}
\author{Chris Nolting}
\affiliation{School of Physics and Astronomy, University of Minnesota, Minneapolis, MN, USA}

\author{T. W. Jones}
\affiliation{School of Physics and Astronomy, University of Minnesota, Minneapolis, MN, USA}

\author{Brian J. O'Neill}
\affiliation{School of Physics and Astronomy, University of Minnesota, Minneapolis, MN, USA}
 
\author{P. J. Mendygral}
\affiliation{Cray Inc., Bloomington, MN, USA}

\begin{abstract}

We report from a study utilizing 3D MHD simulations, including cosmic-ray electrons, of the interactions between radio galaxies (RGs) and dynamically active ICMs. Here we consider interactions involving plane ICM shocks having Mach numbers 2--4 and their normals aligned with steady, active bipolar RG jets penetrating uniform, stationary ICMs. Shock impact disrupts the pre-formed RG jet cocoons into ring vortex structures. Sufficiently strong post-shock winds can stop and even reverse the upwind jet, and strip jets to virtually “naked” states, leaving them without a surrounding cocoon. Strong shock-induced vorticity can also disrupt the downwind jet, so that the ring vortex remnant of the cocoons appears ahead of that jet's visible terminus. 
Magnetic field amplification in the ring vortex can significantly enhance its synchrotron emissions well after the vortex becomes isolated from the RG and its fresh CRe supply. We examine these dynamics and their observable consequences in detail.
\end{abstract}

\section{Introduction}
Galaxy clusters form by way of accretion and occasional mergers. Those behaviors are inherently non-symmetrical and unsteady, which drives strong motions in the diffuse intracluster media (ICMs). Additional ICM ``stirring'' comes, for example, by way of gravitational interactions with nearby halos, galactic winds, and jets. Consequently, ICMs should be dynamic, not static environments. Cluster dynamics include turbulence, sloshing motions, infall flows and shocks \citep[e.g.,][]{Brunetti08,TittleyHenriksen05,ZuHoneRoediger16,Markevitch02, Bruggen12, voit05, Schekochihin09}. Proper characterization of an ICM's dynamical state can reveal much about the cluster's recent history \citep[e.g.,][]{Nagai07, KravtsovBorgani12, Walker19}. On the other hand, ICM motions are quite challenging to measure directly with current instruments.\footnote{The most precise measurements currently available are from \cite{HitomiPerseus16}, which measured a velocity dispersion of $164\pm10$ km/s near the core of the Perseus cluster utilizing the Hitomi SXS instrument.} Some moderately strong ICM shocks with shock normals close to the plane of the sky have been detected in the x-ray band in relatively high density regions \citep[e.g.,][]{Markevitch02, vanWeerenToothbrush16}. Diffuse, non-thermal radio emissions from locally re-accelerated relativistic CRe within ICMs also evidently reveal large-scale dynamical structures including shocks and, probably, turbulence \citep[e.g.,][]{vanWeeren19}. These features are currently the most widely applied signatures of ICM dynamics.

At the same time, radio galaxies (RGs) with non-thermal structures extending 10s--100s of kpc from their active galactic nucleus (AGN) of origin are common throughout cluster volumes, including regions where x-ray emissions are faint \citep[e.g.,][]{padovani16,Garon19}. The properties of these RGs are very sensitive to details of interactions with the ambient ICM. Beyond their source galaxies, RGs are visible largely because of their interactions with ICMs. Because of these interactions, it is important to learn how to use RG properties as ``ICM weather vanes''; i.e., to use RG properties on scales of 10s of kpc and beyond to extract insights about the associated ICM dynamics.

Relative RG--ICM motions and their variations can be particularly telling, since they strongly modify the symmetry of a RG from that provided by the host AGN. Jet bending due to relative motion between the host galaxy and the ICM has been accepted for several decades as the principal cause of the formation of so-called ``head--tail'' RGs \citep[][]{BegelmanReesBlandford,JonesOwen79}. Of course, the existence of a head--tail structure by itself reveals only the presence of relative motion, not the motion of either constituent relative to the cluster center. Other details are necessary to establish the full dynamical picture. In this we should be mindful that because both galaxy motions and bulk ICM motions derive from the same gravitational potential, they can be similar in magnitude. However, galaxy motions are ``particle-like'', while ICMs behave as fluids on these scales.

In recognition of the presence of large scale ICM shocks during merger events, it was pointed out some time ago that the impact of an ICM shock on a RG-formed ICM cavity can transform the cavity into a ``doughnut-like'' ring vortex, while potentially brightening its synchrotron emissions by way of compression and magnetic field amplification \citep[see, e.g.,][for simulation results and possible observed examples]{EnsslinBruggen02,PfrommerJones11}. This topological transformation results from shear induced by the abruptly increased shock speed (and subsequent post-shock flow speed) inside the low density cavity \citep{EnsslinBruggen02}. The same physics has been studied in detail in a laboratory environment using helium bubbles being shocked by air to create vortex rings \citep[e.g.,][]{Ranjan08}. In this work and a companion paper we deal with consequences of the scenario in which a shock interacts with lobes of a RG that contains active jets initially penetrating a stationary medium. The shock passage through the low density lobes of such a RG does lead to formation of vortex rings (that may eventually merge into a single ring). However, the presence of active jets adds considerable richness to the evolution of the impacted RG. Our goals are to understand the underlying physics that controls this evolution and to look for observable dynamical signatures.

There are a growing number of clusters containing ``highly deformed'' RGs whose properties and/or juxtaposition with an x-ray shock might suggest a physical RG--shock encounter \citep[e.g.,][]{WilberNov18,Mandal18}. In some cases the RG morphology and positional relation to the shock suggest that the pre-shock RG had a head--tail structure \citep[e.g.,][]{Bonafede14,Shimwell14,vanWeeren17,Mandal18}.
These shock--radio tail encounters could be useful diagnostics once their evolution is understood and may be relatively easy to spot. However, prior to shock impact, RG tails are complex and heterogeneous structures, which substantially alters and complicates subsequent shock interactions compared to interactions with RG lobes, including those with enclosed jets. 
We will examine the shocked tail problem in a separate paper in preparation \citep{ONeill19a}. We focus first on shock collisions with straight, active, lobed RGs. Active RG jets add significantly to the full evolutionary story, and the character of the jet influence depends on the orientation between the shock normal and the RG jet axis. Fortunately, much of the range in variation can be captured by examining the simple limits in which the jets are either aligned with the impacting shock normal or orthogonal to the shock normal. In this paper we focus on the aligned jet--shock normal case, making reference as needed to consequences when the ideal symmetry is broken. In a companion paper to this one \citep{OrthShockedJet} we conduct an analogous examination of the orthogonal impact case. An additional, complementary paper \citep{ONeill19b} examines in detail the evolution and emissions from AGN jets forming within steady winds; i.e., head--tail formation, including basic dependencies on wind--jet orientation.

\cite{Jones16} emphasized from analytic arguments that one distinctive result of an aligned RG--shock impact is a decreased rate of extension of the forward (upstream) RG jet compared to the extension rate of the downstream jet. For sufficiently strong shocks they suggested that the upstream jet might actually be reversed, leading to essentially one-sided ``head--tail'' RG morphologies. At least one single tail RG has been been identified in a merging cluster that could be a candidate for an aligned shock--RG encounter. Specifically, the ``source C'' in the merging cluster A2256\citep[e.g.,][]{Rottgering94,Owen14} exhibits a narrow (width $\sim1$ kpc even $\ga 100$ kpc projected distance from the AGN), remarkably straight tail extending for a projected distance $\sim 1$ Mpc to the NW of the AGN, with no evident kpc scale structure on the opposite side of the host. The considerable projected length of this tail makes relativistic beaming an unlikely explanation for the asymmetry \citep{deGregory17}, while the large projected length combined with the very narrow width over the large projected distance significantly reduces the odds that it is simply a typical twin-tail RG viewed close to the plane of the two tails (but see \S \ref{subsec:misalign} for more on this issue). We shall see below that, in fact, some basic properties similar to those of source C may be natural outcomes from an aligned RG--shock encounter followed by immersion of the surviving jet in a strong, aligned, post-shock wind. Our purpose here is to explore this basic scenario, not to model that specific object. Our analysis examines the underlying shock--RG dynamics, including evolution of magnetic fields and CRe from the source AGN as well as associated radio synchrotron emissions in simulations motivated by this example.

 The remainder of the paper is organized as follows: Section \ref{sec:interactcartoon} outlines the physical scenario in terms of shock--lobe interactions (\S \ref{subsec:cavities}) and wind--jet interactions (\S \ref{subsec:head}). Section \ref{sec:methods} describes our simulation specifics, including numerical methods (\S \ref{subsec:numerics}) and details of our simulation setups (\S \ref{subsec:Setup}). In section \ref{sec:Discussion} we discuss the results of the simulations, while section \ref{sec:Summary} provides a brief summary of results. 

\section{Outline of Aligned Shock--RG Interaction Dynamics}
\label{sec:interactcartoon}

\begin{figure*}
\centering
\includegraphics[scale=0.7]{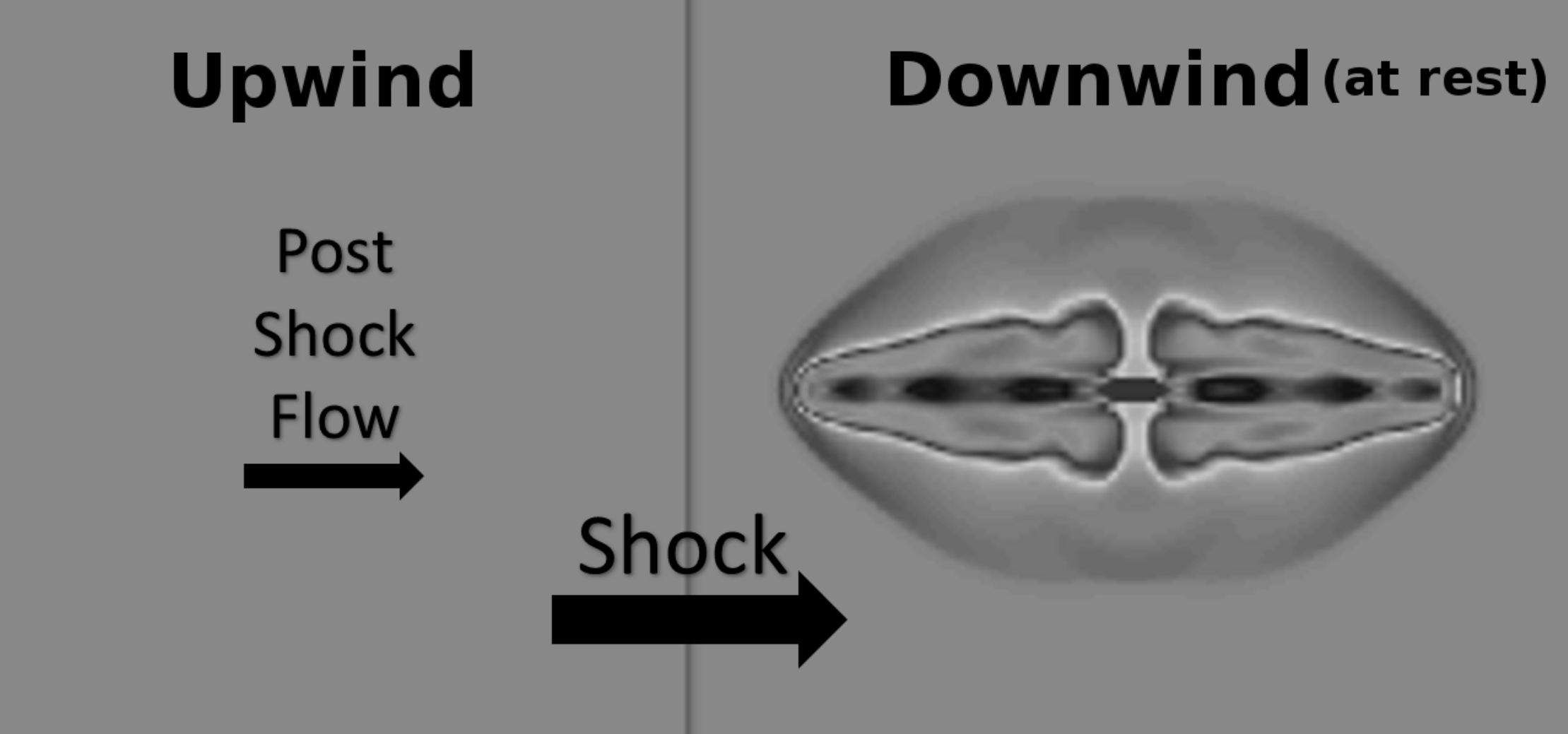}
\caption{Basic geometry of modeled aligned-shock-RG encounters.}
\label{fig:aligned-setup}
\end{figure*}

 The analytic basics of an aligned shock--RG encounter can be outlined rather simply. Our brief discussion largely follows \cite{Jones16}. Readers are referred to that work and references therein for further details. Figure \ref{fig:aligned-setup} illustrates the basic scenario we are addressing in this paper. Although our outline in this section deals strictly with aligned jet--shock encounters, our simulations verify that many aspects of modestly misaligned encounters, aside from obvious symmetry issues, are very similar. Specifically, the treatments of shock propagation through lobes (\S \ref{subsec:cavities}) and jet terminus advancement within a headwind or tailwind (\S \ref{subsec:head}) applying momentum flux in the aligned wind velocity component remain quite useful in understanding misaligned shock--RG encounters. 
 
In the following analysis, all velocities are referenced to the rest frame of the RG-source AGN. We define ``upwind'' to point in the direction the shock is coming from, while ``downwind'' points in the direction the shock is going. Thus, the upwind components of the RG encounter the shock before the downwind components (see Figure \ref{fig:aligned-setup}). In this paper the AGN itself is assumed to be at rest with respect to the unshocked ICM (see \cite{ONeill19b} for alternative choices). For notation clarity we mention that, henceforth, subscripts ``j'' and ``s'' refer to the jet and the shock, while subscripts ``i'' and ``w'' point to the unmodified ICM and the post-shock wind. When combined, as ``ji'' or ``jw'', for example, the notation points to jet properties in the ICM or jet properties in the wind.

Since much of the physics of shock--RG interactions depends on the ICM post-shock pressure, flow velocity and sound speed, we list up front results from the standard Rankine-Hugoniot relations (using $\gamma = 5/3$) connecting the shock Mach number in the ICM, $M_{si} = |v_{si}|/a_i$, and downstream, pre-shock ICM conditions, $P_i$ and $a_i = \sqrt{\gamma P_i/\rho_i}$ to the post-shock, ``wind'' pressure, $P_w$, velocity, $|v_w|$ and sound speed, $a_w$. 
It is often convenient to utilize flow Mach numbers, so we also include the wind Mach number measured with respect to the post-shock pressure and density, $|M_w| = |v_w|/a_w$ (cf. equation \ref{eq:windMach}). Below we treat $v_w$ as a signed velocity component, with $v_w > 0$ corresponding to a headwind as seen by a jet of interest. We have:
%%%%%%%%%%%%%%%%%%%%%
%%%%%%%%%%%%%%%%%%%%%
%%%%%%%%%%%%%%%%%%%%%
\begin{align} 
\label{eq:jump-d}
\rho_w = \frac{4\mathcal{M}_i^2}{\mathcal{M}_i^2+3}\rho_i,\\
\label{eq:jump-p}
P_w = \frac{5M_{si}^2-1}{4}P_i,\\
\label{eq:jump-v}
|v_w| = \frac{3}{4}\frac{M_{si}^2-1}{M_{si}}a_{i},\\
\label{eq:jump-a}
a_w = \frac{\sqrt{(M_{si}^2 + 3)(5 M_{si}^2 - 1)}}{4 M_{si}} a_i,\\
\label{eq:windMach}
|M_w| = \frac{|v_w|}{a_w} = 3 \frac{M_{si}^2 - 1}{\sqrt{(M_{si}^2 + 3)(5M_{si}^2 - 1)}}.
\end{align}
%%%%%%%%%%%%%%%%%%%%%
%%%%%%%%%%%%%%%%%%%%%
%%%%%%%%%%%%%%%%%%%%%
 For the shocks of interest in this study, $2\le M_{si}\le 4$, the post-shock conditions span $2.29 \le \rho_w/\rho_i \le 3.37$, $4.75\le(P_w/P_i)\le 19.75$, $1.13\le |v_w|/a_i\le 7.13$, $1.44\le a_w/a_i\le 1.94$, and $0.23\le |M_w|\le 1.16$.
 %%%%%%%%%%%%%%%%%%%%%%%%%%%%%%%%%%
%%%%%%%%%%%%%%%%%%%%%%%%%%%%%%%%%%
%%%%%%%%%%%%%%%%%%%%%%%%%%%%%%%%%%
%%%%%%%%%%%%%%%%%%%%%%%%%%%%%%%%%%

\subsection{Shock--Lobe Collisions}
\label{subsec:cavities}
Interactions between an ICM shock and a classical, lobed RG can be roughly decomposed into the impulsive impact of the shock discontinuity with its resultant aftermath and the subsequent, longer-term interactions with a post-shock wind flow. The principal modifications to a lobed RG from the shock impact itself come as the shock enters the very low density lobes (i.e., cocoons or cavities). The very low densities of the cavities ($\rho_c \ll \rho_i$, subscript c for cavity) are a direct consequence of their formation by fast, light jets. On the other hand, RG lobes should be in at least rough pressure balance with the unshocked ICM\footnote{In simulations reported here, $\rho_c \la 10^{-2} \rho_i$, with cavity pressure, $P_c \sim P_i$, before ICM shock impact.} ($P_c \sim P_i$). For our needs here only approximate pressure balance matters, and we can similarly ignore any differences in lobe and ICM equations of state. Then, the sound speed in the lobe (cavity), $a_c \sim \sqrt{\rho_i/\rho_c} ~a_i \gg a_i$, where $a_i$ is the sound speed in the unshocked ICM. As the shock enters a low density lobe, the shock velocity increases (so $v_{sc} > a_c > v_{si} = M_{si} a_i$), although, because the lobe plasma is much hotter than the ICM, the internal shock Mach number, $M_{sc} = v_{sc}/a_c$, can be substantially less than the ICM shock Mach number, $M_{si}$; i.e., $M_{sc} < M_{si}$. For $\rho_c \la 10^{-2} \rho_i$ and $M_{si} \sim 2-4$, relevant to our simulations described below, analytic results in \cite{PfrommerJones11} suggest typical expected cavity shock velocities, $v_{sc} \sim 3-5~ v_{si}$ with cavity shock Mach numbers only slightly more than one.

The shock penetrating the cavity pulls with it dense, post-shock ICM at a speed intermediate between the external and internal shock velocities. \cite{PfrommerJones11} find, for conditions similar to those in simulations reported here, a propagation rate for the contact discontinuity (CD) separating the shocked ICM and the lobe plasma of $v_{CD} \sim 1.2 - 1.4~ v_{si}$. Most importantly, the dense ICM inside the original cavity then runs ahead of the surrounding external shock. That generates strong shear along the original boundary between the lobe and the ICM with faster, forward flow inside and the effective return flow around the outside. Since the cross section of the original lobe boundary as seen from the shock plane was roughly circular, the outcome is a toroidal vortex ring tracing that boundary with its axis aligned to the shock normal. (See, e.g., figures \ref{fig:m4synch4}, \ref{fig:aligned-rot60}). 

Our simulations reported below reveal that the hot, low density plasma originally inside the lobes becomes wrapped towards the inside of the vortex torus, surrounded by denser, cooler ICM material that carries most of the kinetic energy in the vortex. Typical for vortex flows, the gas pressure is largest on the outer perimeter, being enhanced in this situation by $\sim \rho_i v_{CD}^2 > M_{si}^2 \rho_i a_i^2$ over the post-shock wind pressure, $P_w$. For the vortex structures generated by $M_{si} \sim 2-4$ shocks, this pressure enhancement can easily exceed $P_i$ by more than an order of magnitude. For the geometries of interest in this paper the circulation around the vortex converges just upwind of where the vortex surrounds the axis of the downwind jet (so slightly closer to the AGN). As we will show, for stronger shocks this can pinch or otherwise disrupt propagation of that jet. This converging flow also leads during this development to significant density enhancement in the post-shock ICM behind the vortex.

\subsection{Jet Propagation in a Headwind or Tailwind}
\label{subsec:head}
The AGN jets themselves primarily respond to conditions in the post-shock flow or wind. For aligned shock--jet encounters in particular, once in the post-shock ICM, the upwind jet finds itself in a high density, high pressure headwind. The downwind jet, if overrun by the ICM shock, will be embedded in a high density, high pressure tailwind. Since, the jet facing upwind into the shock inevitably finds itself in a headwind, we concentrate on that situation, although we lay out the formalism so it applies to either a headwind or a tailwind. In either upwind or downwind situations, except for parameters tuned to match post-shock conditions, the dynamics also apply to more general situations in which an AGN simply finds itself embedded in a wind due to relative AGN--ICM motion. So, with appropriate parameter adjustment the jet behaviors outlined do not require a shock impact. 

The simplest, 1D model for jet terminus (or ``head'') propagation depends only on momentum flux balance at the jet terminus. Then, the adjustment in the advance rate of the terminus for a headwind or a tailwind differs only by a sign in the effective momentum flux for the wind. To compute the propagation velocity of a jet ``head'', $v_h$, we need first the jet thrust, or total jet momentum flux. The fully relativistic momentum flux density of the jet, $T_{mj}$, can be written as $T_{mj} = w_j U_j^2/c^2 + P_j$, where $w_j = e_j + P_j$ is the enthalpy density in the jet, $e_j$ is the energy density including rest mass energy, $P_j$ is the jet pressure and $U_j = \Gamma_j v_j$ is the jet 4-velocity, with $\Gamma_j$ being the jet Lorentz factor. In the non-relativistic limit that applies to our simulations $\Gamma_j \rightarrow 1$ and $w_j \rightarrow \rho_j c^2 + P_j \approx \rho_j c^2$. Then defining the jet sound speed as $a_j = c \sqrt{\partial P_j/\partial e_j} ~_{\overrightarrow{n.r.}}~ \sqrt{\gamma P_j/\rho_j}$, where the final form is the non-relativistic limit. The jet {\underline {internal}} Mach number is $M_j = \Gamma_j v_j/(\Gamma_{s,j} a_j)$, with $\Gamma_{s,j} = 1/\sqrt{1 - (a_j/c)^2}$, so that the total thrust of the jet, $T_{mj} A_j$, can be written as
%%%%%%%%%%%%%%%%%%%%%
%%%%%%%%%%%%%%%%%%%%%
%%%%%%%%%%%%%%%%%%%%%
\begin{align}
 T_{mj}A_j &= \left(w_j\frac{U_j^2}{c^2}+P_j\right)A_j\nonumber\\
           &=(\gamma' M_j^2 +1) P_j A_j
\label{eq:fullthrust}\\
           & _{\overrightarrow{s.s.}} ~\gamma'M_j^2P_jA_j,
\label{eq:jetThrust}
\end{align}
%%%%%%%%%%%%%%%%%%%%%
%%%%%%%%%%%%%%%%%%%%%
%%%%%%%%%%%%%%%%%%%%%
where $A_j$ is the jet cross sectional area and $\gamma' = \left(\Gamma_{s,j}^2a_j^2w_j\right)/(P_j c^2)$. $\gamma' \rightarrow 2$ for an ideal relativistic equation of state while, $\gamma'~_{\overrightarrow{n.r.}} ~\gamma=\frac{5}{3}$ for an ideal monatomic non-relativistic gas. The form in equation \ref{eq:jetThrust} applies when the jet is supersonic. If the jet is supersonic, while both the velocity and equation of state are nonrelativistic, $T_{mj} A_j \approx \gamma M_j^2 P_j A_j = v_j^2 \rho_j A_j$. Although it is common and correct to use $ v_j^2 \rho_j A_j$ to express jet thrust under these circumstances, we prefer in our analysis to use the equivalent $\gamma' M_j^2 P_j A \rightarrow~\gamma M_j^2 P_j A_j$, since for pressure confined jets the local mean jet pressure, $P_j$, should adjust to become comparable to the pressure in the jet surroundings. This form keeps that issue visible. In our simulations the jets are pressure confined, and we find that $P_j$ always adjusts approximately to the surroundings. We note for reference below that if we keep the jet thrust, $T_{mj} A_j$, fixed then $M_j \propto 1/\sqrt{P_j}$ if the jet is supersonic. So, as a supersonic jet propagates from the unshocked ICM into the post-shock wind, the jet Mach number will decrease accordingly. A benefit of tracking the adjusted Mach number is maintaining awareness of whether or not the jet is supersonic at a location of interest.

Our immediate objective is to estimate the propagation velocity of the jet head, $v_h$, with respect to the AGN in the presence of a wind. If the jet head is propagating in an aligned wind of velocity, $v_w$, density, $\rho_w$, and pressure, $P_w$, the total external momentum flux density on the ``nose'' of the jet is
%%%%%%%%%%%%%%%%%%%%%
%%%%%%%%%%%%%%%%%%%%%
%%%%%%%%%%%%%%%%%%%%%
\begin{align}
(v_h + v_w)^2 \rho_w + P_w = \left[\gamma (M_h + M_w)^2) + 1\right] P_w,
\label{eq:jetHeadMomentum}
\end{align}
%%%%%%%%%%%%%%%%%%%%%
%%%%%%%%%%%%%%%%%%%%%
%%%%%%%%%%%%%%%%%%%%%
where $v_w>0$ corresponds to a headwind, while $v_w < 0$ represents a tailwind. In the second form we set $M_h = v_h/a_w$ and $M_w = v_w/a_w$ with $a_w = \sqrt{\gamma P_w/\rho_w}$. Again, $M_w > 0$ refers to a headwind, while $M_w < 0$ is a tailwind. We can estimate $v_h$ or $M_h$ simply by assuming the thrust of the jet in equation \ref{eq:fullthrust} is matched by the external momentum flux distributed over an effective ``head area'', $A_h$, to be estimated empirically in comparison to $A_j$ from simulation results relation to equation \ref{eq:AdvanceRateMachNum} below. The momentum balance relation becomes
%%%%%%%%%%%%%%%%%%%%%
%%%%%%%%%%%%%%%%%%%%%
%%%%%%%%%%%%%%%%%%%%%
\begin{align}
 \left(\gamma'M_j^2 + 1 \right)~P_j A_j &= \left[\gamma (M_h + M_w)^2) + 1\right] P_w A_h,
\label{eq:momentumBalance}
\end{align}
%%%%%%%%%%%%%%%%%%%%%
%%%%%%%%%%%%%%%%%%%%%
%%%%%%%%%%%%%%%%%%%%%
Equation \ref{eq:momentumBalance} can be solved for $M_h$ to give
%%%%%%%%%%%%%%%%%%%%%
%%%%%%%%%%%%%%%%%%%%%
%%%%%%%%%%%%%%%%%%%%%
\begin{align}
 M_{hw} = \frac{v_h}{a_w} \approx - M_w + M_j \sqrt{\frac{\gamma' P_j A_j}{\gamma P_w A_h} + \frac{1}{\gamma M_j^2} \frac{P_j A_j - P_w A_h}{P_w A_h}}\nonumber\\
 _{\overrightarrow{n.r., s.s}}~ - M_w + M_j \sqrt{\frac{A_j}{A_h}},
\label{eq:AdvanceRateMachNum}
\end{align}
%%%%%%%%%%%%%%%%%%%%%
%%%%%%%%%%%%%%%%%%%%%
%%%%%%%%%%%%%%%%%%%%%
where the last line provides the limiting non-relativistic, supersonic form, assuming $P_j = P_w$, while dropping the term $(A_j- A_h)/(\gamma M_j^2A_h)$ inside the square root We should keep in mind that $M_j$ refers to the ``local'', internal Mach number of the jet before its terminal shock. With properly chosen parameters equation \ref{eq:AdvanceRateMachNum} would apply for advancement in any roughly aligned headwind or tailwind. In the absence of a wind, the AGN evolves in a stationary ICM with pressure, $P_i$, and sound speed, $a_i$, and we can express the undisturbed head advance Mach number as $M_{h} = v_{h}/a_i \approx M_{ji} \sqrt{A_j/A_h}$.

Since our focus here is primarily propagation in a post-shock wind, we have, assuming $P_j \sim P_w$,
%%%%%%%%%%%%%%%%%%%%%
%%%%%%%%%%%%%%%%%%%%%
%%%%%%%%%%%%%%%%%%%%%
\begin{align}
 M_{hw} \approx~ - M_w + M_{ji} \sqrt{\frac{P_i A_j}{P_w A_h}},
\label{eq:AdvanceRateModified}
\end{align}
%%%%%%%%%%%%%%%%%%%%%
%%%%%%%%%%%%%%%%%%%%%
%%%%%%%%%%%%%%%%%%%%%
with $M_{ji}$ corresponding to the jet internal Mach number prior to its penetration into the post-shock region. From equation \ref{eq:AdvanceRateModified} we can estimate that in a sufficiently strong headwind the advance of the head reverses, so that the wind drives the jet head back towards its source. In that case there is eventually only a downwind jet; that is, the RG effectively becomes a one-sided jet. Specifically, equation \ref{eq:AdvanceRateModified} predicts that result for $M_w > M_{wc}$, with
\begin{equation}
M_{wc} = M_{ji} \sqrt{\frac{P_i A_j}{P_w A_h}}.
\label{eq:stopjet}
\end{equation}
This dynamics applies only where the jets are exposed to the posited external wind. Near the core of an AGN host galaxy that retains a significant ISM, the upwind jet may very well be isolated from the wind. Such details are beyond the scope of our current study.

\begin{figure*}
\centering
\includegraphics[scale=0.75]{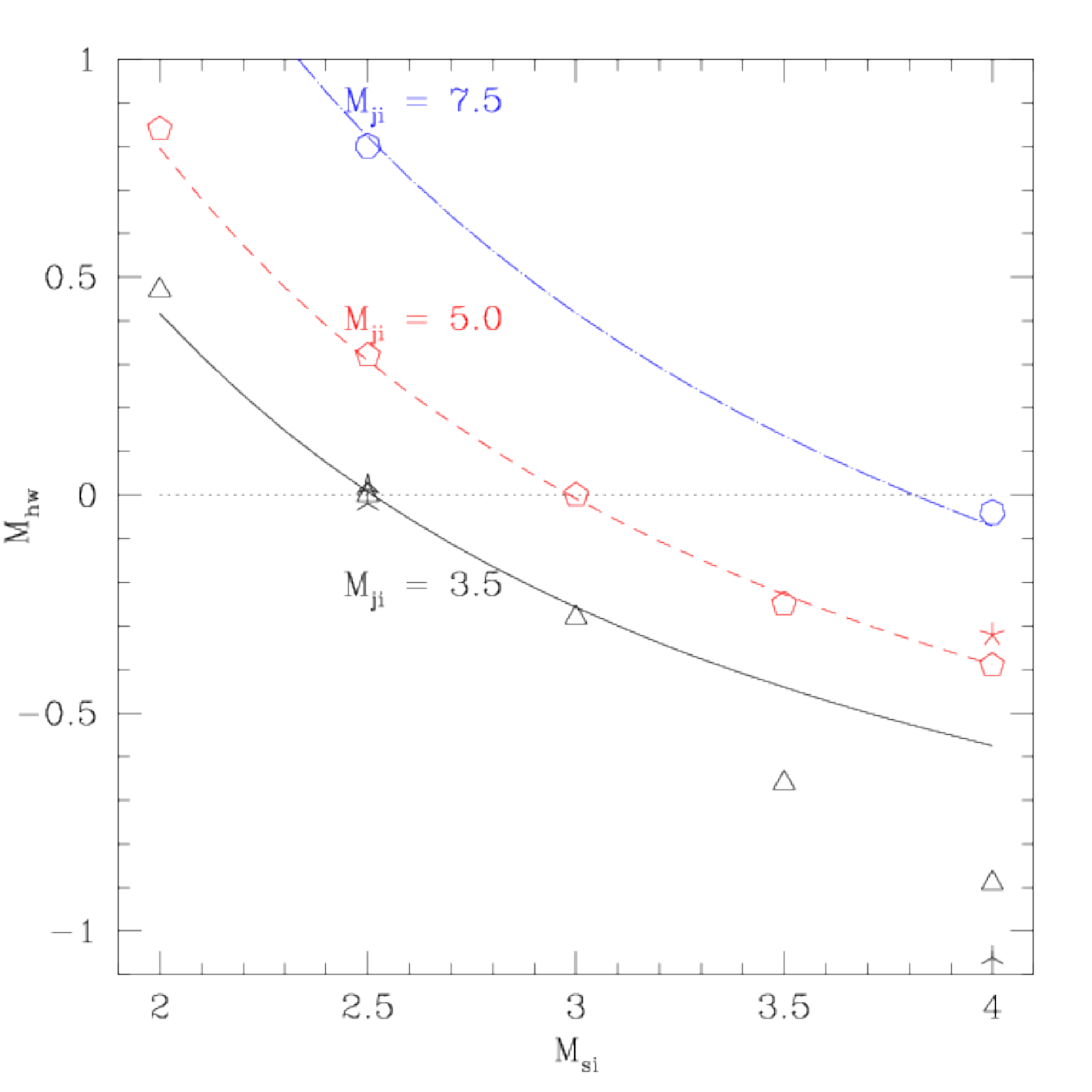}
\caption{Upwind jet head advance Mach number, $M_{hw} = v_h/a_w$, in an aligned post-shock wind vs incident shock Mach number, $M_{si}$. Curves represent equation \ref{eq:AdvanceRateModified} with $\sqrt{A_j/A_h} = 0.7$ and post-shock conditions from equations \ref{eq:jump-p} \& \ref{eq:windMach}, for pre-shock jet Mach numbers $M_{ji} = 3.5 ~\rm{(solid~black)}, 5.0~ \rm{( short-dashed~red)}, 7.5 ~\rm{(long-dashed~blue)}$. Points with colors matching the curves are empirically measured results from simulations with matching parameters (see section \ref{sec:methods}). Point symmetries approximately reflect $M_{ji}$ for the associated simulation. Values below the dotted horizontal line at $M_{hw} = 0$ indicate the upwind jet head advance is reversed by the wind ($v_{hw} < 0$). }
\label{fig:vhead_test}
\end{figure*}

The preceding simple analysis is illustrated in Figure \ref{fig:vhead_test}. Curves plot $M_{hw}$ from equation \ref{eq:AdvanceRateModified} (setting $\sqrt{A_j/A_h} = 0.7$). Values empirically determined from 3D MHD parameter-matching simulations detailed in section \ref{sec:methods} have been added for comparison. The dotted horizontal line at $M_{hw} = 0$ corresponds to the condition for $M_{wc}$ in equation \ref{eq:stopjet}. The comparison simulated jets were initially in pressure balance with the pre-shock ICM ($P_j = P_i$), had pre-shock Mach numbers, $M_{ji} = 3.5, 5.0, 7.5$, matching the curves. They had a range of magnetizations and density contrasts with the ICM as explained in \S \ref{subsec:Setup}. With the exception of two $M_{ji} = 3.5$ comparison simulations with $M_{si} = 3.5$ and $M_{si} = 4$, the simulations confirm predictions of equation \ref{eq:AdvanceRateModified}. In both of the exceptional cases the high post-shock pressure made the jet subsonic ($M_j < 1$), so that equation \ref{eq:AdvanceRateModified} provides a poor approximation to $M_{hw}$. A better estimate would be $M_{hw} \sim -M_w +\sqrt{A_j/(\gamma A_h)}$, which is less than found from equation \ref{eq:AdvanceRateModified} provided the trivial constraint on the pre-shock conditions, $\gamma M_{ji}^2 > 1$, is satisfied.

%%%%%%%%%%%%%%%%%%%%%%%%%%%%%%%%%%
%%%%%%%%%%%%%%%%%%%%%%%%%%%%%%%%%%
%%%%%%%%%%%%%%%%%%%%%%%%%%%%%%%%%%
%%%%%%%%%%%%%%%%%%%%%%%%%%%%%%%%%%

It is straightforward to estimate the shock strengths that lead the jets to become subsonic in the post-shock flow. In particular, using equation \ref{eq:jump-p} and assuming the post-shock jet pressure satisfies $P_j \sim P_w$, we conclude for $\gamma = 5/3$ that the jets in the post-shock flow become subsonic if
%%%%%%%%%%%%%%%%%%%%%
%%%%%%%%%%%%%%%%%%%%%
%%%%%%%%%%%%%%%%%%%%%
\begin{equation}
\frac{M_{si}}{M_{ji}} > \sqrt{ \frac{4}{5} + \frac{1}{5 M_{ji}^2} }.%~ ~_{\overrightarrow{M_{ji}\gg 1}}~~ 0.9.
\label{eq:subsonicjet}
\end{equation}
%%%%%%%%%%%%%%%%%%%%%
%%%%%%%%%%%%%%%%%%%%%
%%%%%%%%%%%%%%%%%%%%%

Although our focus in this paper is on winds that are aligned with the AGN jets, we do in \S \ref{subsec:misalign} briefly examine the situation in which a headwind effectively stops a jet that is moderately misaligned with the wind. In that case it is more appropriate to describe the dynamical influence of the wind on the upwind AGN jet as a deflection or bending of the jet downwind rather than stopping the jet. In \cite{ONeill19b}, we explore generally how winds bend jets, but it is a convenient test of our analysis later in this paper to evaluate the condition identified in equation \ref{eq:stopjet} in terms of the deflection such a wind would produce if it were misaligned. In particular we note that equation \ref{eq:stopjet} can be rewritten as $v_w^2 \rho_w = v_j^2 \rho_j (A_j/A_h)$. Then, if we recall the ``classic'' relation for the ratio of a jet radius to the bending radius of curvature, $\ell_b$, in a transverse wind \citep[][]{BegelmanReesBlandford, JonesOwen79}, $r_j /\ell_b \sim v_w^2 \rho_w /(v_j^2 \rho_j)$, we see immediately that equation \ref{eq:stopjet} corresponds roughly to the condition that a transverse wind would bend a propagating jet on a scale comparable to the jet radius. Thus, we would expect a wind satisfying equation \ref{eq:stopjet} impacting an oblique jet to very sharply deflect that jet. We confirm that behavior in a simulation outlined in \S \ref{subsec:misalign} and also in our analysis in \cite{ONeill19b}.

\section{Simulation Specifics}
\label{sec:methods}

\subsection{Numerical Methods}
\label{subsec:numerics}
The simulations reported here were carried out employing the Eulerian WOMBAT ideal 3D nonrelativistic MHD code described in \cite{PeteThesis} on a uniform, Cartesian grid using an adiabatic equation of state with $\gamma = 5/3$. The simulations utilized the 2nd order TVD algorithm with constrained magnetic field transport as in \cite{Ryu98}. Simulation setup specifics follow in \S \ref{subsec:Setup}.

The simulations also incorporated a conservative Eulerian Fokker-Planck solver for transport of the cosmic ray electron (CRe) distribution, $f(p)$, employing the ``coarse grained momentum volume transport'' (CGMV) algorithm introduced by \cite{JonesKang05}. The CGMV method provides an economical way to track CRe populations over a wide range of particle momenta (energies), including relevant physical processes. We outline here only a few essentials of the method, referring readers to \cite{JonesKang05} and references therein for further details. In the current simulations the isotropic CRe momentum distribution, $(n_{CRe} \propto \int p^2 f(p) dp = \int p^3 f(p) d\ln p$), was continuous in the range $10 \la p/(m_e c)\la 1.7\times 10^5$ (so, energies 5 MeV $\la E_{CRe} \la$ 90 GeV) with uniform logarithmic momentum bins, $1\le k\le 8$. 
Within a given bin, $k$, the momentum distribution was assumed to be a power law, $f(p) \propto p^{-q_k}$, although $q_k$ varied between bins. This distribution was evolved according to the history of a given CRe population. We included adiabatic, as well as radiative (synchrotron and inverse Compton) energy changes outside of shocks, along with test-particle diffusive shock (re)acceleration (DSA) at any shocks encountered. We did not include CRe energy losses from Coulomb collisions. For the thermal plasma densities relevant to these simulations collisional losses are subdominant to inverse Compton/synchrotron losses unless $E_{CRe} \la 200$ MeV \citep[e.g.,][]{Sarazin99}, whereas synchrotron emissions examined in this study generally involve $E_{CRe} \ga 1$ GeV. For $E_{CRe} \la 200$ MeV energy loss time scales generally exceed the simulation times in any case.

 DSA was implemented at shock passage by setting $q_{k,out} = \min(q_{k,in},3\sigma /(\sigma - 1))$ immediately post-shock, where $\sigma$ is the code-evaluated compression ratio of the shock. 
This simple treatment is appropriate in the CRe energy range covered, since typical DSA acceleration times to those energies are much less than a typical time step in the simulations ($\Delta t \ga 10^4$ yr). Our simulations primarily target lower luminosity, ``FRI'', RGs in which the relativistic plasma on multi-kpc scales is energetically subdominant \citep[e.g.][]{CrostonHardcastle14}. Accordingly, our CRe populations were passive. The total CRe number density, $n_{CRe}$, was arbitrary, since it had no impact on dynamical evolution. Consequently, we can compute meaningful synchrotron brightness, polarization and spectral distributions from our simulations, but synchrotron intensity normalizations that we present are arbitrary. 
We did not here include spatial or momentum diffusion in our CRe, so neglected 2nd order, turbulent particle reacceleration. 

Except for a negligible background CRe population included in the ICM to avoid numerical singularity problems in the CGMV algorithm, all the CRe in our simulations were injected onto the computational domain as part of the AGN jet generation process (outlined in the following subsection). 
At injection from the AGN source, the CRe momentum distribution was a power law with $q = q_0 = 4.2$, over the full momentum range. This translates into a synchrotron spectral index, $\alpha = 0.6$ ($F_{\nu} \propto \nu^{-\alpha})$ using the conventional relation for power laws. Our synchrotron emissions reported here are computed using the actual $f(p)$ over the momentum range specified above along with the full synchrotron emissivity kernel for isotropic electrons, given a local magnetic field $\vec{B}$ \citep[e.g.,][]{BlumenthalGould70B}.

\subsection{Simulation Setup}
\label{subsec:Setup}

%%%%%%-Start Table 1%%%%%%%%%%
\begin{deluxetable}{ccccccccc}
  \tabletypesize{\footnotesize}
   \tablewidth{0pt}
 \tablecaption{Simulation Jet  Parameters\label{table:jetparams}}
   \tablehead{
   \colhead{Run} & \colhead{$M_{ji}$} & \colhead{$\rho_i/\rho_j$} & \colhead{$v_{ji}$} &\colhead{$\beta_{pj}$} & $B_0$ & \colhead{$\theta_j$*} & \colhead{$x_{jc}$} & \colhead{$r_j$} \\
   \colhead{}  & \colhead{($v_{ji}/a_{ji}$)} & \colhead{} & \colhead{($10^4$ km/sec)} & \colhead{($P_i/P_{B0}$)} & \colhead{($\mu$G)}  & \colhead{(degrees) } & \colhead{(kpc)} & \colhead{(kpc)}
    }
 \startdata
 \hline
 \bf{J3S2}  & 3.5 & $10^2$ & 2.3 & 75 & 2.1 & 0 & -80 & 3.0\\
 \hline
 J3S22 & 3.5& $10^2$  & 2.3 & 75 & 2.1 & 15 & -288 & 3.0\\ 
 J3S25a & 3.5 & $10^2$& 2.3 & $10^3$ & 0.6 & 0 & -146  & 4.0\\
 J3S25b  & 3.5 & $10^3$ & 7.3 & $10^3$ & 0.6 & 0 & -146 & 4.0\\
 J3S25c  & 3.5 & $10^2$  & 2.3 & 10 & 5.8 & 0 & -146 & 4.0\\
 J3S3  & 3.5 & $10^2$ & 2.3 &$10^3$  & 0.6 & 0 & -146 & 4.0\\
 J3S35 & 3.5 & $10^2$ & 2.3& $10^3$ & 0.6 & 0& -123 & 4.0\\
 \hline
 \bf{J3S4a} & 3.5 & $10^2$  & 2.3 &  75 & 2.1 & 0 & -208 & 3.0\\
 \hline
 J3S4b & 3.5 & $10^3$ & 7.3 & $10^3$ & 0.6 & 0 & -78 & 4.0\\
 J5S2 & 5.0 & $10^2$  & 3.3 & $10^3$ & 0.6 & 0 & -123 & 4.0\\
 J5S25 & 5.0 & $10^2$  & 3.3 & $10^3$ & 0.6 & 0 & -146 & 4.0\\
 J5S3  & 5.0& $10^2$  & 3.3 & $10^3$  & 0.6 & 0 & -123 & 4.0\\
 J5S35 & 5.0 &$10^2$  & 3.3 & $10^3$  & 0.6 & 0 & -123 & 4.0\\
 J5S4a  & 5.0 & $10^2$ & 3.3 & $10^3$  & 0.6 & 0 & -78 & 4.0\\
 J5S4b  & 5.0& $10^3$ & 3.3 & $10^3$  & 0.6& 0 & -78 & 4.0\\
 J7S25  & 7.5 & $10^2$  & 5.0 & $10^3$ & 0.6 & 0 & -123 & 4.0\\
 J7S4 & 7.5& $10^2$  & 5.0 &$10^3$ & 0.6  & 0 & -45 & 4.0\\ 
 \enddata
 \tablecomments{All simulations had steady, bipolar jets launched from coordinate $x_{jc}$ (with $y_{jc} = z_{jc} = 0$) in the x-y plane at $t=0$ into a static, uniform and unmagnetized ICM that had density, $\rho_i = 5\times 10^{-27}$ g/cm$^3$, and pressure, $P_i = 1.33 \times 10^{-11}$ dyne/cm$^2$, so sound speed, $a_i = 667$ km/sec. $\theta_j$ measures the angle between the jet axis and shock normal. At launch jets had (gas) pressure, $P_j = P_i$, so internal sound speed, $a_{ji} = a_i  \sqrt{\rho_i/\rho_j}$.  In all reported simulations, $\Delta x = \Delta y = \Delta z = 0.5$ kpc.}
 \label{table:tab1}
 \end{deluxetable}
 %%%%%%-End -Table 1%%%%%%%%%%

%%%%%%-Start Table 2%%%%%%%%%%
\begin{deluxetable}{ccccccc}
  \tabletypesize{\footnotesize}
   \tablewidth{0pt}
 \tablecaption{Simulation Shock \& Domain  Parameters\label{table:shockparams}}
   \tablehead{
   \colhead{Run} & \colhead{$M_{si}$} & \colhead{$P_{w}/P_i$} & \colhead{$v_{w}$} & \colhead{$x_{domain}$} &\colhead{$y_{domain}$} & \colhead{$z_{domain}$} \\
   \colhead{} & \colhead{} & \colhead{} & \colhead{($10^3$ km/sec)} & \colhead{(kpc)} & \colhead{(kpc)} & \colhead{(kpc)} 
    }
 \startdata
 \hline
\bf{ J3S2} & 2.0 & 4.75 & 0.75 & $\pm$ 320 & $\pm$ 149 & $\pm$ 149 \\
\hline
J3S22 & 2.25 & 6.08 & 0.90 & $\pm$ 464 & $\pm$ 240 & $\pm$ 144 \\ 
 J3S25 & 2.5 & 7.56 & 1.05 & $\pm$ 224 & $\pm$ 50 & $\pm$ 50\\
 J3S25b & 2.5 & 7.56 & 1.05 & $\pm$ 224 & $\pm$ 50 & $\pm$ 50 \\
J3S25c  & 2.5 & 7.56 & 1.05& $\pm$ 224 & $\pm$ 50 & $\pm$ 50 \\
 J3S3 & 3.0 & 11.0 & 1.33 & $\pm$ 224& $\pm$ 50 & $\pm$ 50\\
J3S35  & 3.5 & 15.1 & 1.61 & $\pm$ 224& $\pm$ 50 & $\pm$ 50\\
\hline
\bf{J3S4a}  & 4.0 & 19.8 & 1.88 & $\pm$ 448 & $\pm$ 149 & $\pm$ 149 \\
\hline
J3S4b  & 4.0 & 19.8 & 1.88 & $\pm$ 224 & $\pm$ 50 & $\pm$ 50 \\
 J5S2  & 2.0 & 4.75 & 0.75 & $\pm$ 224 & $\pm$ 50 & $\pm$ 50 \\
 J5S25 & 2.5 & 6.08 & 1.05 & $\pm$ 224& $\pm$ 50 & $\pm$ 50 \\
J5S3  & 3.0 & 11.0 & 1.33 & $\pm$ 224 & $\pm$ 50 & $\pm$ 50 \\
J5S35  & 3.5 & 15.1 & 1.61 & $\pm$ 224 & $\pm$ 50 & $\pm$ 50 \\
J5S4a  & 4.0 & 19.8 & 1.81 & $\pm$ 224 & $\pm$ 50& $\pm$ 50 \\
J5S4b & 4.0 & 19.8 & 1.81 & $\pm$ 224 & $\pm$ 50 & $\pm$ 50 \\
J7S25 & 2.5 & 7.56 & 1.05 & $\pm$ 224& $\pm$ 50 & $\pm$ 50 \\
J7S4 & 2.0 & 19.8 & 1.88 & $\pm$ 224 & $\pm$ 50 & $\pm$ 50 \\ 
 \enddata
 \tablecomments{In all simulations, shocks entered the domain from the $-x$ boundary propagating in the $+\hat x$ direction. The post-shock, ``wind'' velocity, $v_w$, is computed from equation \ref{eq:jump-v}, while $P_w$ is from equation \ref{eq:jump-p}. In all reported simulations, $\Delta x = \Delta y = \Delta z = 0.5$ kpc. }
 \label{table:tab2}
 \end{deluxetable}
 %%%%%%-End -Table 2%%%%%%%%%%

For this study we carried out sixteen 3D MHD simulations involving bipolar AGN jets. With one exception the jet axis aligned with the normal of an incident plane ICM shock. In one simulation the jet axis was offset by 15 degrees from the shock normal. Simulation parameters are outlined in Tables \ref{table:jetparams} (jet properties) and \ref{table:shockparams} (shock and ICM properties). These simulations cover a range of internal jet Mach numbers from $3.5 < M_{ji} < 7.5$ and external ICM shock Mach numbers of $2 < M_{si} < 4$. The jet and shock Mach numbers provide the basis for labeling the various simulations. For example, the simulation $\bf J3S2$ involved jet Mach number, $M_{ji}=3.5$ and shock Mach number, $M_{si}=2$ (keeping only one digit of the Mach number in the label). In several cases where we considered multiple shock Mach numbers between integer values we added a second digit to the shock Mach number in the label. For instance, simulation $\bf J3S2$ involved $M_{si} = 2.0$, while simulation $\bf J3S22$ involved $M_{si} = 2.25$. In three situations where we varied either the jet density or magnetic field strength (always maintaining the jet gas pressure, $P_j = P_i$ and jet Mach number, $M_{ji}$ at launch), we added a letter ``a'', ``b'' or ``c'' at the end of the label. For example, simulations $\bf J3S4a$ and $\bf J3S4b$ both involved $M_{ji} = 3.5$ and $M_{si} = 4$, but had $\beta_{pj}$ values of $75$ and $10^3$, respectively. Our detailed analysis in section \ref{sec:Discussion} focuses primarily on two aligned shock--jet simulations, {\bf J3S2} and {\bf J3S4a}, that roughly span the range of dynamical behaviors we obtained. Specifics of those simulation properties are emphasized in Tables \ref{table:tab1} and \ref{table:tab2}. Section \ref{sec:Discussion} also includes a brief comparison of {\bf J3S22}, which breaks the alignment symmetry of the other simulations. The simulation domains, while varying in size according to the needs of a given dynamical situation, were all rectangular prisms with the origin at the domain center ($x = y = z = 0$). All grid boundaries in the simulations were open, except for inflow conditions fixed at the left $x$ boundary to drive the planar ICM shock in the $+x$ direction.

All the simulated AGN jets reported here were steady, as well as collimated and bi-directional at launch, beginning by definition at $t=0$. They were launched out of a cylinder whose axis was in the x-y plane and was at rest on the grid, centered at the middle of the y-z plane (y = z = 0). In all but one of the simulations the jet axis was along the grid x axis to align with the external shock propagation. The central x coordinate for the jet launch cylinder, $x_{jc}$, was adjusted for each simulation according to dynamical parameters to delay boundary interactions associated with the jets as long as practical (specifics given for each simulation in Table \ref{table:shockparams}). 

Jet launch cylinders were 24 grid cells long, with similar jets emerging from each end. They had radius either 6 or 8 cells, as indicated in Table \ref{table:jetparams}. The launch cylinders were surrounded by a coaxial cylindrical collar two grid cells thick providing transition from conditions maintained inside the launch cylinder to ambient conditions outside. Our launched jets were dynamically composed of non-relativistic (thermal) weakly magnetized ($\beta_{pj}>>1$, as defined below) plasma, since we have in mind FRI jets that on these scales have entrained substantial ISM plasma from their galaxies of origin \citep[e.g.,][]{Croston18}. As noted above, we did include a passive, relativistic CRe population in order to model nonthermal emissions. To trace directly the distribution of plasma injected onto the grid through the jet launch cylinder, we advected a passive ``jet mass fraction'' scalar that was set to unity for material entering the grid via the jet launch cylinder, but initialized at zero elsewhere. 

The simulated jets were magnetized inside the launch cylinder by a uniform, poloidal electric current whose sign matched the jet velocity at a given location. The result was a toroidal magnetic field inside the launch cylinder, $\vec{B} = \pm B_0 (r/r_j)\hat{\phi}$, where $r_j$ is the outer jet radius, and $\hat{\phi}$ is the azimuthal unit vector relative to the launch cylinder axis. The nominal jet magnetic field strength, $B_0 = B(r_j)$, was set by fixing the plasma $\beta_{pj} = P_{i}/P_B(r_j)$ at the launch cylinder perimeter, $r_j$, where $P_B(r_j) = B_0^2/(8\pi)$ is the magnetic pressure at $r_j$. For the simulations reported here $\beta_{pj}$ ranged between $10$ and $10^3$, as listed in Table \ref{table:jetparams}. There was a poloidal return current in the launch cylinder transition collar, so that the net electric current along the jet launch cylinder vanished everywhere. The ambient ICM, both undisturbed and post-shock were initially unmagnetized in all the simulations reported here. Of course, any ambient ICM that mixed with jet plasma became magnetized and also carried CRe originating in the jets. We acknowledge that real ICM plasmas are magnetized, but our approach maximizes our ability to understand how AGN magnetic fields evolve in response to the dynamical scenarios under study. 

The emergent jet density, $\rho_j$, gas pressure, $P_j = P_i$, and velocity, $v_j = M_{ji} a_{ji}$, were uniform across the ends of the launch cylinder, where $a^2_{ji} = \gamma P_i/\rho_j = \rho_i/\rho_j~ a^2_i$. The simulations reported here mostly set $\rho_i/\rho_j = 10^2$ but we also included two cases with $\rho_i/\rho_j = 10^3$.
The $\rho_i/\rho_j = 10^2$ contrast is a compromise that allows for a light jet while keeping the simulation cost reasonable. Although detailed lobe morphology varies with $\rho_i/\rho_j$, the dynamical behaviors at the center of this study; namely, the propagation of the RG head and the interaction of the jets with the post-shock wind were insensitive to this ratio (see, e.g., Figure \ref{fig:vhead_test}).
Due to these density ratios, either $a_{ji} = 10 a_i$ or $a_{ji} \approx 31.6 a_i$. The launched jet Mach number, $M_{ji} = 3.5$ in our detailed analysis simulations {\bf J3S2} and {\bf J3S4a}, but in other simulations ranged between 3.5 and 7.5 as indicated in Table \ref{table:jetparams}. All the simulations reported here included plane shocks propagating through the stationary, homogeneous ICM along the x axis, entering from the left boundary. ICM shock Mach numbers, $M_{si}$, ranged between 2 and 4, as indicated in Table \ref{table:shockparams}. 

Up to this point in our presentation we have provided no explicit physical length, time, density or pressure scales for the simulations. That is because, except for radiative energy losses by the passive CRe, the simulations are scale free, meaning none of the dynamical outcomes in our simulations depend on those choices. In particular, since the fluids are ideal, adiabatic and non-relativistic,{\emph { all the dynamical behaviors can be fully referenced to dimensionless parameters, such as Mach number, $\beta_{pj}$ and ambient to jet density ratio and normalized by a characteristic length, such as $r_j$, a characteristic pressure, such as $P_i$, and a characteristic velocity, such as $a_i$.}} We do this purposefully, so that readers can choose the scales that best suit their specific purpose. However, since we do include radiative behaviors of the CRe at specified frequencies in order to identify observable behaviors, representation of the CRe populations and their associated emissions require us to identify explicitly length, velocity and magnetic field scales. On the latter point we emphasize that the magnetic field and pressure scales are uniquely linked by $\beta_{pj}$. In addition, since CRe radiative losses include inverse Compton scattering of CMB photons, and that is a function of redshift, emissions analysis requires us also to specify redshift. For the simulations reported here we set the redshift to be z = 0.2, for which the inverse Compton, radiative cooling time, $\tau_{rad} \sim 60~(10 GeV/E_{CRe})$ Myr, where $E_{CRe} = p m_e c$ is the CRe energy. For $z \rightarrow 0$ the inverse Compton cooling time at a given CRe energy is approximately doubled. The angle-averaged synchrotron energy loss rate matches the inverse Compton rate for $B \approx 3.25 (1+z)^2\mu G$, corresponding approximately to $4.7\mu G$ at $z=0.2$. In the source frame the synchrotron critical frequency can be expressed as $ \nu_c \approx 1.7~{\rm GHz}~(E_{CRe}/10 GeV)^2 B_{\mu G}$. We found that our dynamical results were relatively insensitive to $\beta_{pj}$ within the range we explored, so with appropriate rescalings of emission frequency and/or radiative lifetimes (while properly accounting for inverse Compton cooling), our synchrotron emission images could be adjusted realatively simply for different $\beta_{pj}$ choices.

 In practice, we set our simulation length scale so that a grid cell spanned $\Delta x = \Delta y = \Delta z = 0.5$ kpc. Our jet launch cylinder length was then 6 kpc from the center to each end. At launch the jets had $r_j = 3$ kpc or $r_j = 4$ kpc, depending on the number of cells radially spanning the launch cylinder. The characteristic ICM pressure and sound speed were $P_i = (4/3)\times 10^{-11}$ dyne/cm$^2$ and $a_i = (2/3)\times 10^3$ km/sec $\approx (2/3)$ kpc/Myr, respectively, so that $\rho_i = 5\times 10^{-27}$ g/cm$^3$. All our simulations included uniform pre-shock ICMs with these properties. While of course this is a dramatic simplification from real cluster environments, this choice helps us to identify the physical behaviors resulting from the dynamical scenario under study, and not due to complications from additional structure in the medium. With these characteristic scalings the jet magnetic field strength parameter, $B_0 = \sqrt{8\pi P_i/\beta_{pj}} \approx 18 \mu{\rm G}/\sqrt{\beta_{pj}}$, so $B_0$ ranged from about 0.6 $\mu$G ($\beta_{pj} = 10^3$) to about 6 $\mu$G ($\beta_{pj} = 10$). Consistent with our scalings in the previous paragraph CRe synchrotron energy losses are subdominant to inverse Compton energy losses at $z = 0.2$, unless $B \ga 4.7 \mu$G. Thus, except in flow regions where the magnetic field became significantly amplified by stretching, non-adiabatic CRe energy losses were dominated by inverse Compton cooling, which is uniform in time on the scales relevant to these simulations. If our simulated objects were moved to redshift, $z = 0$, the minimum magnetic field required for synchrotron losses to dominate inverse Compton losses would drop to $3.3\mu$G. CRe re-acceleration at shocks via DSA, as well as adiabatic energy changes were included, although only a few of the strongest shocks that developed in these simulations were strong enough to significantly influence the ``observed'' synchrotron emissions.

\section{Discussion}
\label{sec:Discussion}
We now examine and compare three of the simulations from Tables \ref{table:jetparams} and \ref{table:shockparams}. Our primary focus is on two of these; namely, {\bf J3S4a} and {\bf J3S2}, which span the behaviors of our aligned jet ($\theta_j = 0$) simulations. In order to illustrate the consequences of moderate misalignment between the AGN jets and the ICM shock normal, we also briefly discuss the ``misaligned'' simulation {\bf J3S22} with $\theta_j = 15^{\degree}$. All three simulations involved AGN jets with Mach number, $M_{ji} = 3.5$, jet mass density, $\rho_j = 10^{-2} \rho_i$ and characteristic magnetic field strength, $B_0 = 2.1~\mu$Gauss. The only significant difference between the two aligned simulations, {\bf J3S4a} and {\bf J3S2}, was the ICM shock Mach number. {\bf J3S4a} involved $M_{si} = 4.0$, while in {\bf J3S2} the shock had $M_{si} = 2.0$.
The shock in {\bf J3S22} had $M_{si} = 2.25$, which we had estimated in advance as capable of just stopping the advance of the upwind AGN jet, taking into account that some of the jet momentum was in the $\hat{y}$ direction, as a $M_{si} = 2.5$ shock would be required for an aligned jet with the same properties (cf. equation \ref{eq:stopjet}, figure \ref{fig:vhead_test}).

In all three simulations slightly more than $50$ Myr of steady, undisturbed AGN evolution passed before shock first contact. The properties of the RGs at initial shock contact were, accordingly, quite similar. In particular, during this pre-shock evolution each jet head advanced into the ICM with Mach number, $M_{ho}= v_{ho}/a_i \approx 2.5$, consistent with equation \ref{eq:AdvanceRateMachNum} using $M_w = 0$ and $\sqrt{A_j/A_h} \approx 0.7$ ($A_h/A_j \approx 2$). At first shock contact each of the RG lobes had a length $\sim 90$ kpc. We note that although the detailed shapes of RG lobes in our simulations depended on the density ratio\footnote{The pre-shock lobes in simulations {\bf J3S25b} and {\bf J3S4b} with $\rho_j/\rho_i = 10^{-3}$ were fatter than the lobes in simulations with $\rho_j/\rho_i = 10^{-2}$, such as {\bf J3S4a} and {\bf J3S2}.}, $\rho_j/\rho_i$, the velocity of the jet heads, and so the lengths of the RG lobes, did not depend on $\rho_j/\rho_i$. 

Because of their near axial symmetry, the post-shock evolution of both {\bf J3S4a} and {\bf J3S2} were simpler than that of {\bf J3S22}, though all three were qualitatively consistent with the dynamical scenario outlined in section \ref{sec:interactcartoon}. In particular, the incident shocks propagated rather quickly through the low density RG lobes and generated recognizable vortex ring structures. The upwind jet in {\bf J3S4a} was strongly reversed in the post-shock wind, and as predicted for these parameters in \S \ref{subsec:head}, the AGN jets in {\bf J3S4a} actually became subsonic in this flow\footnote {We verified in other simulations that an equivalent wind without shock impact led to essentially the same jet propagation properties, including its subsonic character.}. The upwind jet in the misaligned case, {\bf J3S22}, was sharply bent into the downwind direction, although it remained supersonic. The upwind jet in {\bf J3S2} continued to advance into the post-shock wind at a speed, $v_{hw}$, consistent with equation \ref{eq:AdvanceRateModified}.

\subsection{Simulation J3S4a: $\theta_j =0$, $M_{ji} = 3.5$, $M_{si} = 4.0$}
\label{subsec:M4}
\begin{figure*}
\centering
\includegraphics[width=15cm]{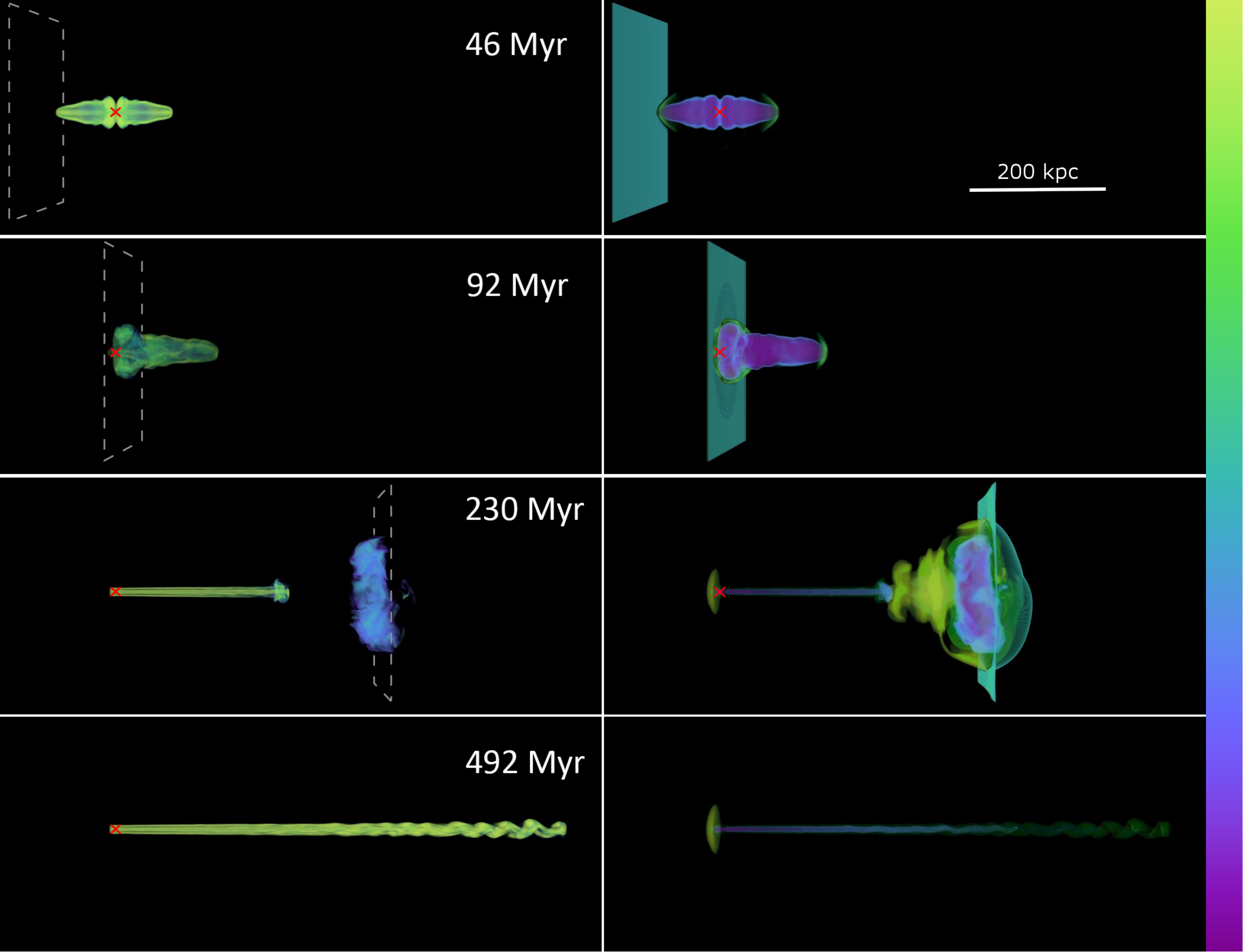}
\caption{ Volume renderings of the {\bf J3S4a} simulation at four times increasing top to bottom. The shock normal and jet axis are in the viewing plane. Shock impact on the RG began soon after the top snapshot. Left: Jet mass fraction ($>$ 30\% visible); Right: Log mass density spanning 3 decades in $\rho$, with key dynamical structures highlighted. The external ICM shock is outlined in dashed gray in the left images, while it can be directly observed in the right images. Colors in all images follow the ``CubeYF'' colormap with ``yellow'' high and ``purple'' low. The AGN location is marked in each image by a red ``x''. }
\label{fig:m4evolve}
\end{figure*}

The dynamical RG evolution of the {\bf J3S4a} RG--shock encounter is outlined in Figure \ref{fig:m4evolve}. The figure presents four snapshots of the volume-rendered\footnote{As viewed along the $\hat{z}$ axis at a distance roughly 1.5 Mpc from the AGN.} jet mass fraction tracer (left panels) and mass density (right panels) at: (1) $t =46$ Myr, just prior to RG--shock first contact (refer to Figure \ref{fig:aligned-setup} for the geometry); (2) $t = 92$ Myr, after the post-shock flow reversed the upwind AGN jet; (3) $t = 230$ Myr, after the shock-generated vortex ring became fully developed and, as discussed in \S \ref{subsec:cavities}, had just pinched off, so truncated, the downwind jet; and (4) $t = 492$ Myr, after the ICM shock and the vortex ring (advected in the post-shock wind) left the simulation box to the right. The structures remaining in the computation domain at the latest time were very close to what would result from a pure head-on wind with the same properties, but absent any shock impact. We note that for $t \ga 92$ Myr the RG itself became essentially a one-sided jet outside the jet launch cylinder. This outcome should follow so long as the jet and wind combine to satisfy the dynamics reflected in equation \ref{eq:stopjet}.

%%%%%%%%%%%%%%%%%%%%%%%%%%%%%%%%%%
%%%%%%%%%%%%%%%%%%%%%%%%%%%%%%%%%%
%%%%%%%%%%%%%%%%%%%%%%%%%%%%%%%%%%
%%%%%%%%%%%%%%%%%%%%%%%%%%%%%%%%%%

At $t = 92$ Myr, the accelerated internal shock had propagated fully through both lobes, while the much slower external shock (outlined in dashed gray on the left of figure \ref{fig:m4evolve}, and emphasized in the density images on the right by enhancing the emissivity of the shock in the external medium) had just passed the position of the AGN (marked with an ``x''). 
The external ICM shock in this simulation propagated substantially faster than the pre-shock downwind RG head ($M_{si} = 4$ vs $ M_{ho} \approx 2.5$). Consequently by $t = 230$ Myr the external ICM shock had reached almost as far as the shock-modified RG then extended downstream. It was located just to the right (downstream) of the vortex ring visible in the mass fraction rendering. The convex shock structure visible to the right of the external ICM shock in the density rendering highlights the location of the accelerated, internal shock after it exited the RG downwind lobe ahead of the external shock.
The vortex-ring-induced high mass density region mentioned in \S \ref{subsec:cavities} and surrounding the jet just to the left (upstream) of the vortex ring is also obvious at $t = 230$ Myr. Thermal Bremsstrahlung images we constructed (not shown) indicated an enhanced x-ray flux of $\sim22\%$ from this region. This suggests such high density features developed from shock-induced vortex structures could be detectable during this stage in deep x-ray ICM images.  We leave exploration of this effect and its observational signatures to future work.

\begin{figure*}
\centering
\includegraphics[width=12cm]{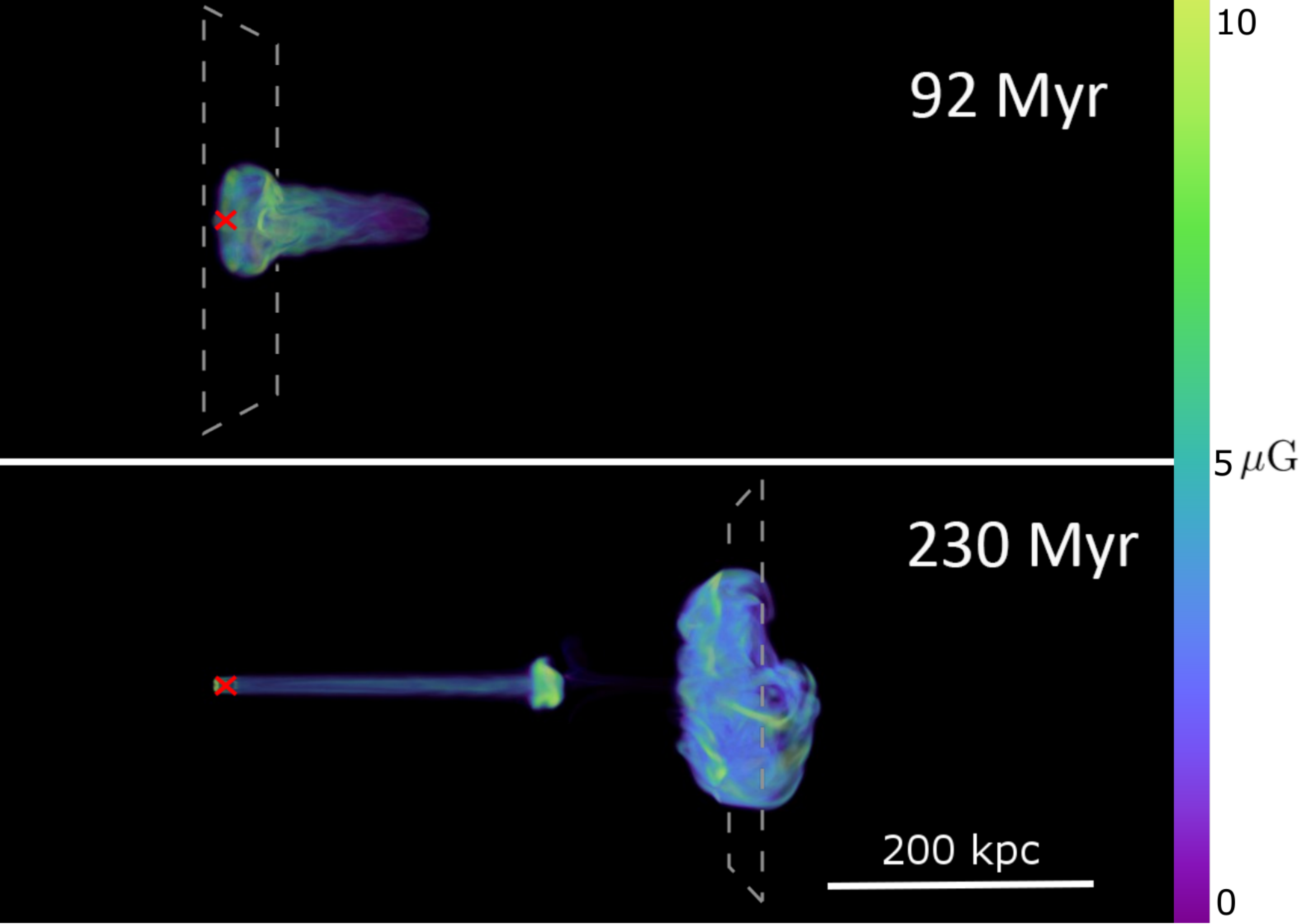}
\caption{ Volume rendering of magnetic field intensity in the {\bf J3S4a} simulation at the two intermediate times and the same perspective as in Figure \ref{fig:m4evolve}.
The emergent AGN jet carries a peak 2 $\mu$G field into an unmagnetized ICM. Fields visible in the image span roughly 1 $\mu$G to 10 $\mu$G. As in figure \ref{fig:m4evolve}, the location of the external ICM shock is outlined in gray.
}
\label{fig:m4Bfield}
\end{figure*}

\begin{figure*}
\includegraphics[width=17.5cm]{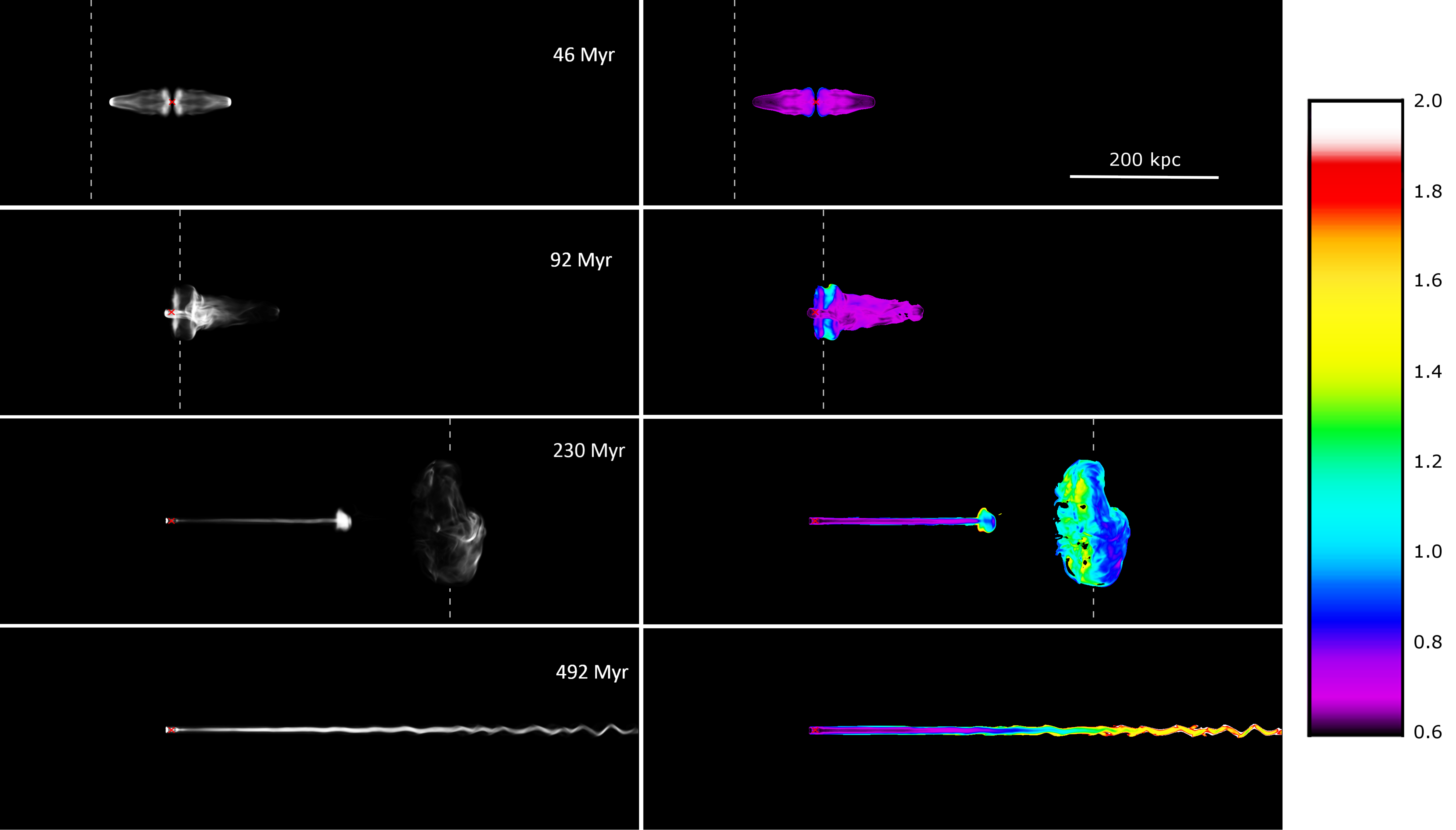}
\caption{Synchrotron images from {\bf J3S4a} at the times in Figure \ref{fig:m4evolve}. Resolution is 0.5 kpc. The AGN jet axis and shock normal are in the plane of the sky. Left: 150 MHz intensity with the brightest pixel in each image approximately the highest intensity excluding the jet launch cylinder at that time. Right: 150/600 MHz spectral index, $\alpha_{150/600}$, for regions above $0.5$\% of the peak intensity at 150 MHz. Spectral index scale on the far right. At launch the jet spectral index was $\alpha = 0.6$. The location of the external ICM shock is denoted by a dashed gray line.}
\label{fig:m4synch4}
\end{figure*}

\begin{figure}
\centering
\includegraphics[width=14cm]{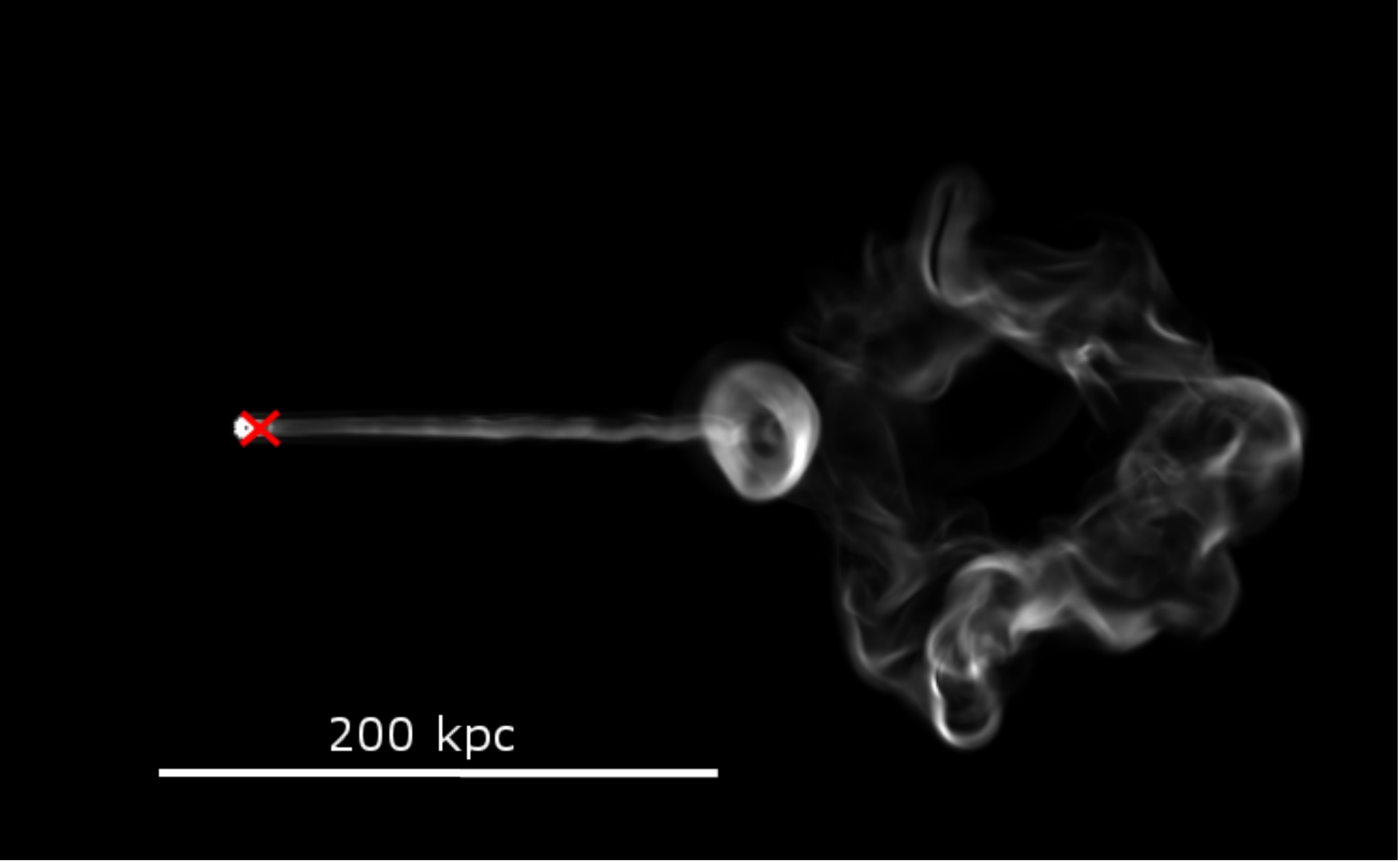}
\caption{150 MHz synchrotron image from {\bf J3S4a} at $t= 302$ Myr with the downwind jet axis projected away from the observer and 30 degrees from the line of sight in order to reveal the ring nature of the vortex ring produced by the shock impact and the ``proto-lobe'' at the terminus of the truncated downwind jet.}
\label{fig:aligned-rot60}
\end{figure}

The jet mass fraction rendering at $t = 230$ Myr clearly reveals the vortex ring. It is also obvious from the mass fraction view at this time that the right-facing, downwind jet, having recently been pinched off by circulation of the vortex ring, no longer continued to the extreme downwind RG extension. However, since the AGN was still active, the truncated downwind jet terminus visible just to the left of the vortex ring continued to drive downwind (to the right). But the now subsonic jet, rather than forming a new RG lobe extending well back towards the AGN, was surrounded at its terminus by a more limited ``proto-lobe'' of AGN plasma. The strong post-shock wind prevented that plasma from extending a substantial distance back towards the AGN. That is, it never formed a classic RG lobe. It is, nonetheless, a distinct backflow feature from the AGN jet.
We point out below that the magnetic field in this proto-lobe was relatively strong and that, consequently, it was remarkably radio bright (see Figures \ref{fig:m4Bfield}, \ref{fig:m4synch4}, \ref{fig:aligned-rot60}).

Since pinching or otherwise disrupting the downwind AGN jet by the vortex circulation produced distinctive features, it is appropriate to outline the essential conditions required for this development. As mentioned in \S \ref{subsec:cavities}, the excess ``on-axis'' pressure towards the perimeter of the ring vortex, $P_{ex}\sim \rho_i v_{CD}^2 \ga M_{si}^2 \rho_i a_i^2$, where $v_{CD} > v_{si}$ is the speed at which the contact discontinuity between RG lobe plasma and post-shock ICM plasma propagated through a lobe. Disruption of the propagating jet by such a pressure imbalance across it would require roughly that $P_{ex} \ga \rho_j v_j^2$. This leads to a very rough condition estimate for disruption of the jet by the vortex flow, $M_{si} \ga M_{ji}$. That is, when the ICM shock Mach number is comparable to or exceeds the Mach number of the AGN jet in the pre-shock flow, the chances are high that the downwind jet may be disrupted by the flow of the vortex ring. In {\bf J3S4a} that condition is easily satisfied.

%%%%%%%%%%%%%%%%%%%%%%%%%%%%%%%%%%
%%%%%%%%%%%%%%%%%%%%%%%%%%%%%%%%%%
%%%%%%%%%%%%%%%%%%%%%%%%%%%%%%%%%%
%%%%%%%%%%%%%%%%%%%%%%%%%%%%%%%%%%

Additional insights into the dynamical evolution of {\bf J3S4a} are evident in Figure \ref{fig:m4Bfield}, which presents volume renderings of the magnetic field strength at the middle two times in Figure \ref{fig:m4evolve}. Recall that the ambient ICM was unmagnetized in this simulation, while the peak magnetic field in the emergent jets was $B_0 \approx 2~\mu$G. The ``naked'' jet visible in the lower, later image in the figure exhibited fields close to those emergent strengths. In fact, the magnetic field in the jet retained to a large degree its original toroidal form to very large distances. However, there were substantial regions in the vortex ring, both as it was first forming (cf. $t = 92$ Myr in the figure) and after it had separated from the AGN (cf. $t = 230$ Myr in the figure) where the field strengths reached and even exceed $10~\mu$G in filaments evidently spanning the torus. The same is true in the proto-lobe at the end of the truncated downwind jet visible at $t = 230$ Myr.

The dynamical link to the development of these strong fields is that those structures contain strong vortical motions that have stretched and amplified the magnetic field coming originally from the jets. In fact, the RMS magnetic field strength in the vortex ring continued to increase until the structure exited the computational domain just after $t \sim 302$ Myr. The peak magnetic field in the proto-lobe also increased with time, although the RMS field did not change significantly. There were, additionally, magnetic field regions in the RG lobes amplified by shear prior to the shock encounter, although the strong shock-induced shear was much more effective in this regard. Since significant ICM plasma was entrained into the lobes as well as the vortex ring, had our ICM been magnetized, the fields in all these regions would likely have been stronger than we see here \citep[see, e.g.][]{Tregillis01,ONeill05,HuarteEspinosa11}.

%%%%%%%%%%%%%%%%%%%%%%%%%%%%%%%%%%
%%%%%%%%%%%%%%%%%%%%%%%%%%%%%%%%%%
%%%%%%%%%%%%%%%%%%%%%%%%%%%%%%%%%%
%%%%%%%%%%%%%%%%%%%%%%%%%%%%%%%%%%

%%%%%%%%%%%%%%%%%%%%%%%%%%%%%%%%%%
%%%%%%%%%%%%%%%%%%%%%%%%%%%%%%%%%%
%%%%%%%%%%%%%%%%%%%%%%%%%%%%%%%%%
%%%%%%%%%%%%%%%%%%%%%%%%%%%%%%%%%

Figure \ref{fig:m4synch4} presents $0.5$ kpc resolution radio synchrotron images of the {\bf J3S4a} structures at the same times as in Figure \ref{fig:m4evolve}. The plane of the sky includes the AGN jets and the ICM shock normal. Each image is integrated from the synchrotron emissivity along the line of sight. The left panels show the synchrotron brightness (arbitrary units) at 150 MHz, while the right panels show associated 150 MHz/600 MHz spectral index maps for the emission down to 0.5\% of the peak 150 MHz intensity in each image. As a reminder of the ``doughnut'', toroidal topology of the vortex ring structure, we also show in Figure \ref{fig:aligned-rot60} the 150 MHz image of {\bf J3S4a} at $t = 302$ Myr with the downwind jet axis and the shock normal now projected $60^{\degree}$ from the plane of the sky.

Two key things are immediately evident from the intensity images. First, the synchrotron surface brightness in this frequency band for this source traces very well regions of strong magnetic field amplification (cf. Figure \ref{fig:m4Bfield}). At the earliest time shown ($t = 46$ Myr), the field amplification in the unshocked lobes mentioned above is particularly visible in the central regions near the jet source. Second, at late times, when it was the only dynamical structure linked to the AGN, the downwind jet remained moderately bright at these frequencies over distances approaching Mpc scales. At the latest time shown ($t = 492$ Myr), when only the jet remained in the simulation domain, the jet surface brightness actually peaked more than 100 kpc from its source. We note that the synchrotron image at this time is at least qualitatively reminiscent of the one sided tail, ``Source C'' in A2256 both in scale and form \citep{Owen14}. In considering this comparison it is worth keeping in mind that many properties of the simulated AGN at this time, after the vortex ring has been advected away, are predominantly controlled by the jet interactions with the surrounding wind. The primary role of the incident shock at this stage was its generation of the strong, high pressure wind. But, any dynamics capable of setting up such a relative wind would lead to rather similar behaviors.

The spectral index images in Figure \ref{fig:m4synch4} add useful insights to the surface brightness behaviors for {\bf J3S4a}. 
In particular, $\alpha_{150/600}$ remains close to the source value, $\alpha = 0.6$, in regions still ``in direct communication with'' the AGN, until the last image at $t = 492$ Myr. At the three earlier times, ($t = 46$ Myr, $t = 92$ Myr and $t = 230$ Myr) only plasma that was or had been wrapped into the vortex ring (i.e., isolated from the jets) showed clear evidence of spectral steepening (aging) in this band. At $t = 492$ Myr, when only the downwind jet remained in the domain, the jet spectrum remained reasonably flat over much of its length, with $\alpha_{150/600} \la 1$ over a distance out to roughly 1/2 Mpc from the source. Near where the jet exited the computational domain, however, $\alpha_{150/600}\ga 1.5$. It should be noted that CRe aging in this tail was influenced at a given distance from the AGN by the fact that the relatively low Mach number of the launched jet in this simulation, $M_{ji} = 3.5$, combined with the high pressure in the strong, post-shock wind caused the jet to become subsonic relatively slower than at launch. A jet with an initially higher Mach number would have remained supersonic to the end (cf. \S \ref{subsec:head}).

In general the modest aging behaviors just described are consistent with the fact that in much of the visibly emitting volume in this source, $B < 5 \mu$G. There, CRe aging was dominated by inverse Compton cooling (recall that z = 0.2). At the same time, in most of the synchrotron-visible regions, $B\ge 1 \mu$G, so that synchrotron emissions at frequencies $\la 1$ GHz came from CRe with $E_{CRe} \la ~{\rm few}$ GeV. Under these circumstances CRe radiative lifetimes, $\tau_{rad} \ga 100$ My. As emphasized above, the magnetic field strengths in the vortex ring and in the proto-lobe feature sometimes reached or exceeded $B \sim 10 \mu$G, which, not only made those structures bright, but also reduced $\tau_{rad}$ to something closer to 30 Myr for CRe emitting near 1 GHz. That explains why in Figure \ref{fig:m4synch4} the spectra in those features are relatively steeper than in some other bright regions.

\begin{figure}
\centering
\includegraphics[width=8cm]{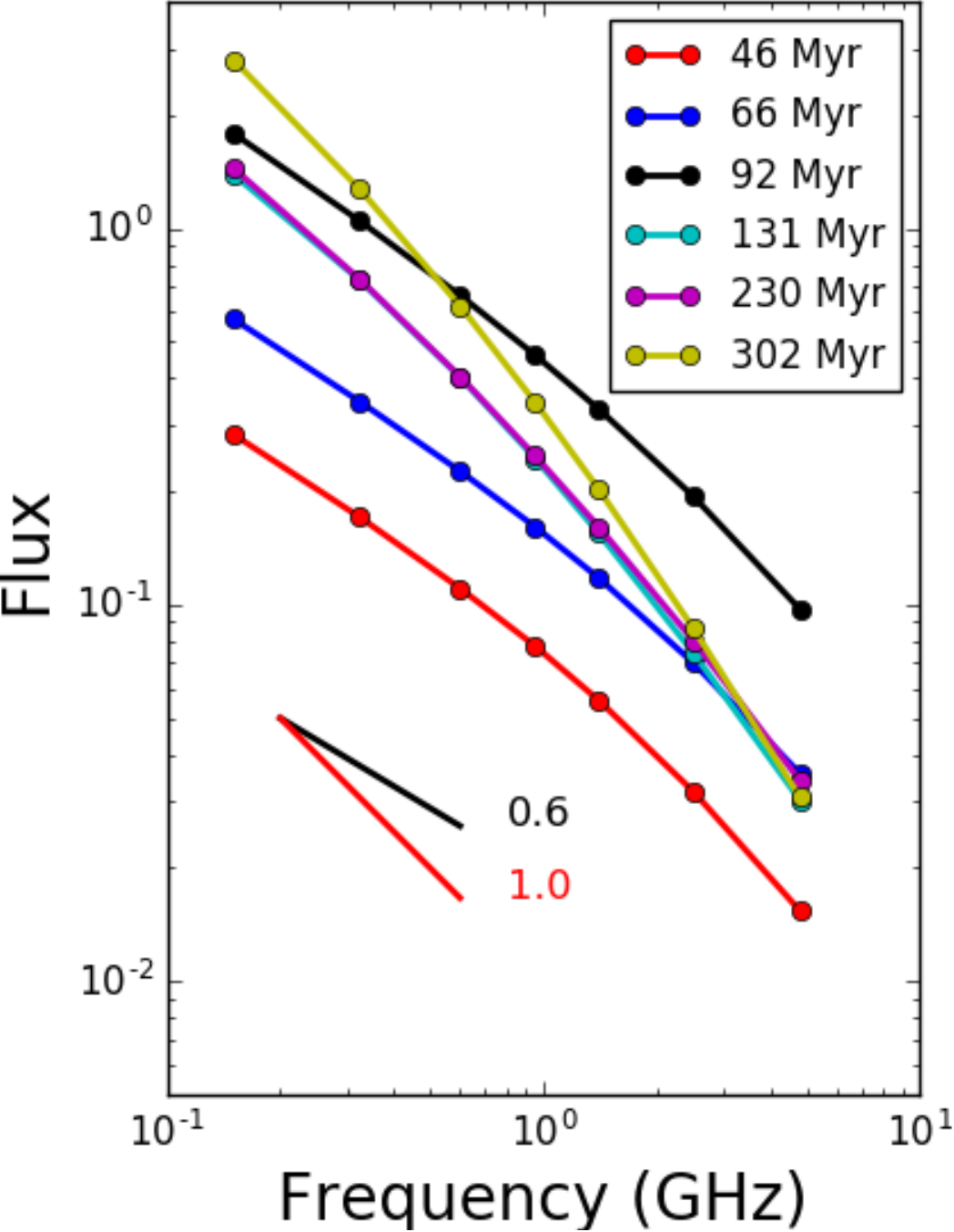}
\caption{ Evolution of the full, integrated {\bf J3S4a} synchrotron spectrum before the vortex ring left the grid. Power law spectra with $\alpha = 0.6$ and $\alpha = 1.0$ are shown for comparison.}
\label{fig:spectralEvolution}
\end{figure}

The behaviors just outlined facilitate interpretation of the integrated synchrotron spectrum and its evolution as shown at multiple times in Figure \ref{fig:spectralEvolution} between $t = 46$ Myr and $t = 302$ Myr. We see in particular that prior to formation of the vortex ring between $t = 92$ Myr and $t = 131$ Myr the integrated luminosity grows at a faster-than-linear pace, due in combination to continued, fresh CRe injection from the AGN and moderate magnetic field amplification. At the same time the spectrum maintains close to the injected spectral index, $\alpha = 0.6$ at low frequencies and exhibits very modest shape evolution due to aging. Once plasma in the vortex ring is disconnected from the AGN prior to $t = 131$ Myr, the integrated spectrum starts to age significantly. 

%%%%%%%%%%%%%%%%%%%%%%%%%
%%%%%%%%%%%%%%%%%%%%%%%%%
%%%%%%%%%%%%%%%%%%%%%%%%%
%%%%%%%%%%%%%%%%%%%%%%%%%

\begin{figure}
\centering
\includegraphics[width=5.6cm]{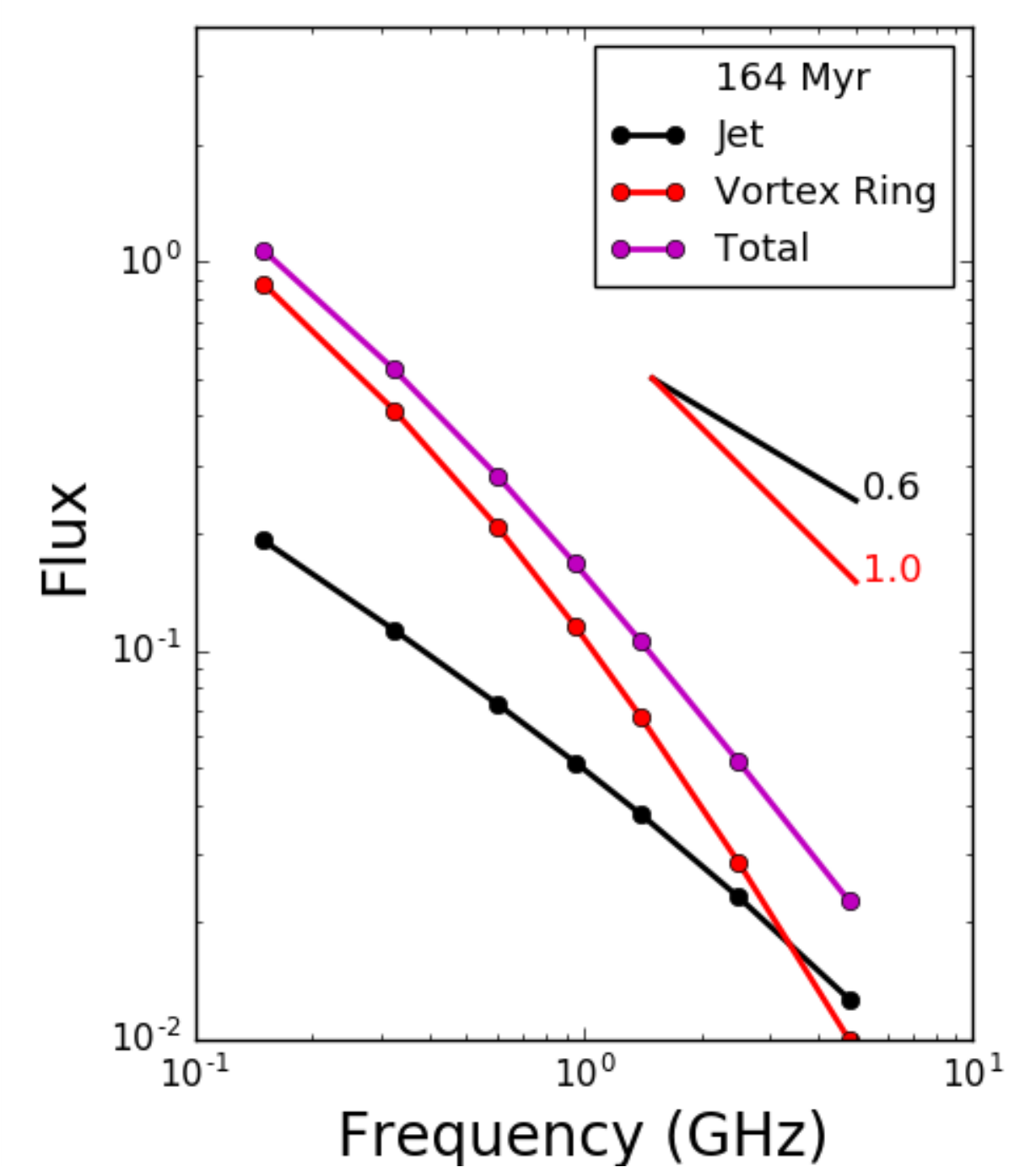}\includegraphics[width=5.6cm]{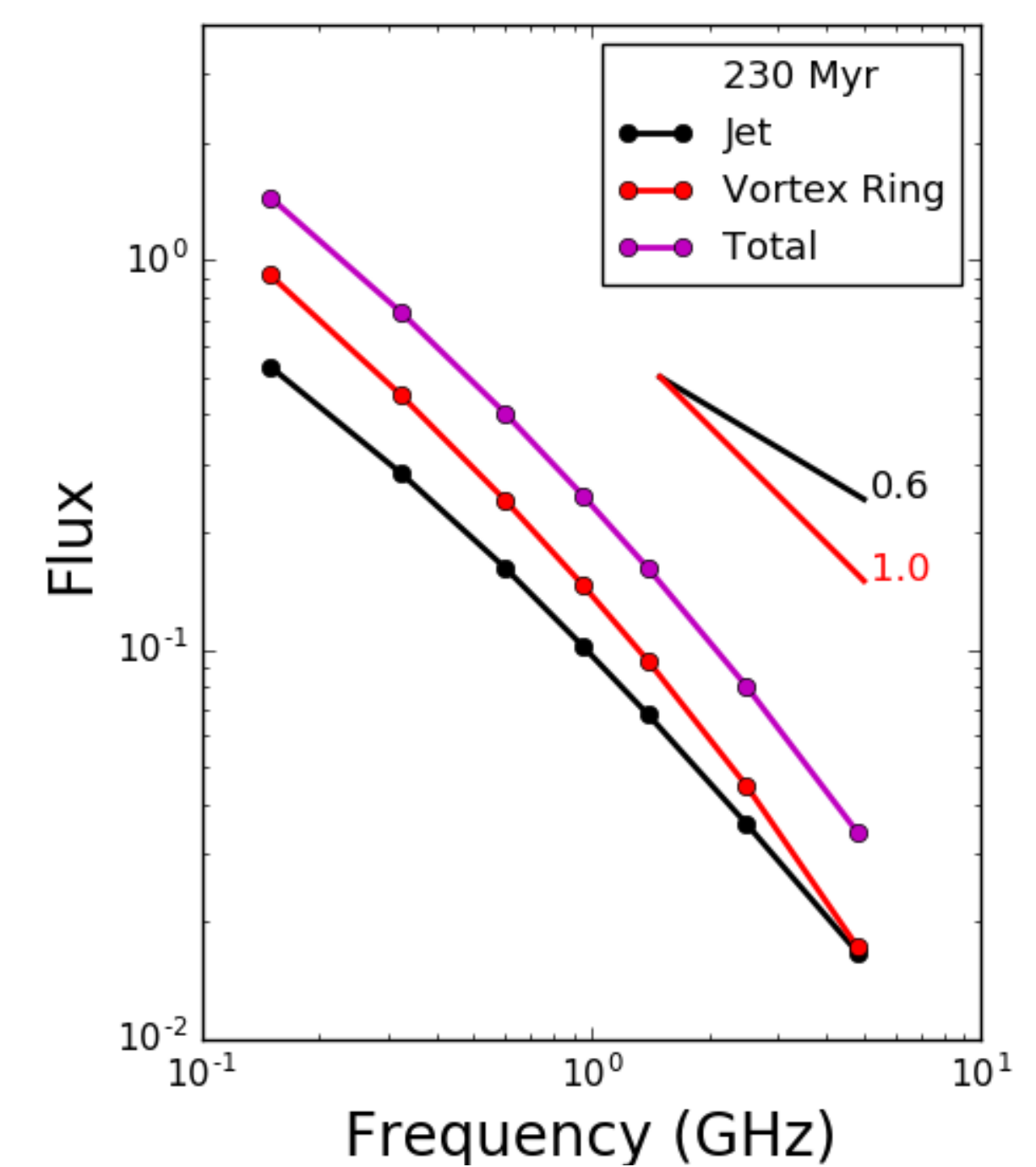}\includegraphics[width=5.6cm]{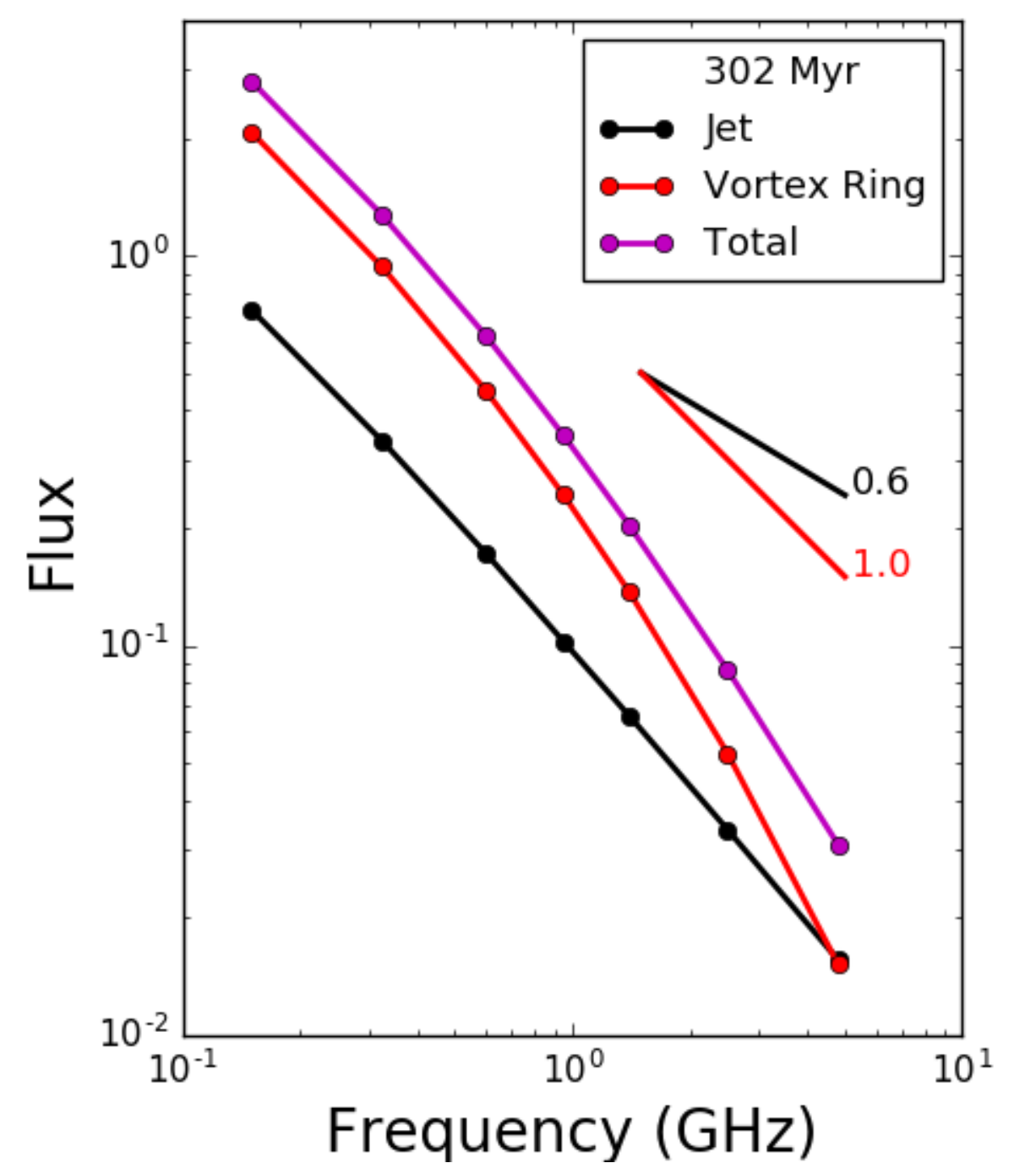}
\caption{Late evolution of the {\bf J3S4a} synchrotron spectra isolating contributions from the disconnected vortex ring.}
\label{fig:m4spec-decomp}
\end{figure}

The importance of the vortex ring to the integrated properties is apparent in Figure \ref{fig:m4spec-decomp}, where the total flux at three times ($t = 164$ Myr, $t = 230$ Myr and $t = 302$ Myr) has been separated into contributions from the vortex ring and everything else (the ``jet'' in the labels). It is clear that the spectrum of the vortex ring became significantly steeper and more convex over time compared to the emissions coming from regions still in communication with the AGN. It is also notable that, except at high frequencies, the total flux from the vortex ring is generally greater than the remainder of the source. That is a consequence of the continued magnetic field amplification in the vortex ring, which persisted in the simulation until the vortex ring left the computational domain. We did not include turbulent CRe re-acceleration in this simulation, but that would likely have enhanced the radio luminosity of the vortex ring even further, while reducing spectral aging. Details of those rather uncertain processes are beyond the scope of this study.

%%%%%%%%%%%%%%%%%%%%%%%%%
%%%%%%%%%%%%%%%%%%%%%%%%%
%%%%%%%%%%%%%%%%%%%%%%%%%
%%%%%%%%%%%%%%%%%%%%%%%%%

\subsection{Simulation J3S2: $\theta_j =0$, $M_{ji} = 3.5$, $M_{si} = 2.0$}
\label{subsec:M2}
%%%%%%%%%%%%%%%%%%%%%%%%%%%%
%%%%%%%%%%%%%%%%%%%%%%%%%%%%
\begin{figure}
\centering
\includegraphics[width=14cm]{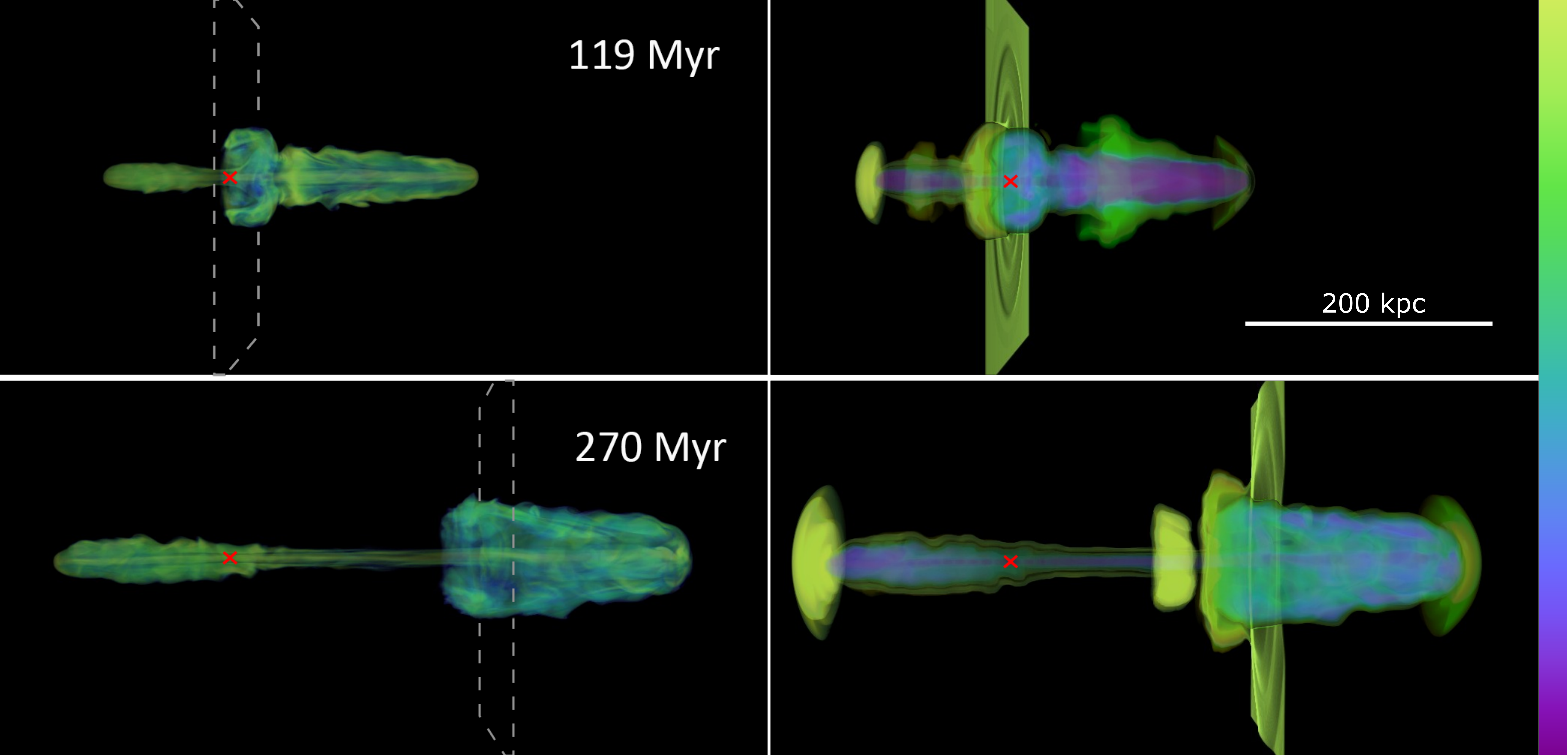}
\caption{ Volume renderings of the {\bf J3S2} simulation at two times following shock impact. Orientations the same as Figure \ref{fig:m4evolve}. Left: Jet mass fraction ($>$ 30\% visible); Right: Log mass density spanning 3 decades in $\rho$, with key dynamical features highlighted. The ICM shock is visible at both times. AGN location is marked by a red ``x''. As in figure \ref{fig:m4evolve}, the location of the external ICM shock is outlined in dashed gray on the left, and directly observable in the density on the right.}
\label{fig:m2evolve}
\end{figure}
%%%%%%%%%%%%%%%%%%%%%%%%%%%%
%%%%%%%%%%%%%%%%%%%%%%%%%%%%

Figure \ref{fig:m2evolve} presents volume renderings of the jet mass fraction (left) and log mass density (right) from the {\bf J3S2} simulation at times $t = 119$ Myr and $t = 270$ Myr. Recall that the only significant difference between the {\bf J3S2} and the {\bf J3S4a} simulations was the strength of the incident ICM shock along with its associated post-shock wind. Although both shock impacts have obvious consequences, the weaker shock in this case ($M_{si} = 2$) has less immediately obvious dynamical influence on the RG than the $M_{si} = 4$ shock in {\bf J3S4a}. In both cases shock interaction with the RG began just after $t = 50$ Myr. 
Qualitatively, the dynamical state at $t = 119$ Myr in Figure \ref{fig:m2evolve} is similar to the $t = 92$ Myr state after the stronger shock impact shown in Figure \ref{fig:m4evolve}. Noting that an ``x'' marks the location of the AGN in each case, however, it is already evident that the weaker shock in {\bf J3S2} has not completely stopped the advance of the upwind AGN jet. This is in contrast to the situation in {\bf J3S4a}, where the upwind jet was reversed. The rate of upwind head advance in {\bf J3S2} is, on the other hand, consistent with equation \ref{eq:AdvanceRateModified}, as shown in Figure \ref{fig:vhead_test}. Indeed it is obvious that the upwind jet at $t = 270$ Myr in the lower panel in Figure \ref{fig:m2evolve} has extended its length since the earlier time in the upper panel. A compact AGN plasma cocoon (RG lobe) also remains around the upwind jet.
%%%%%%%%%%%%%%%%%%%%%%%%%%%%
%%%%%%%%%%%%%%%%%%%%%%%%%%%%
\begin{figure}
\centering
\includegraphics[width=12cm]{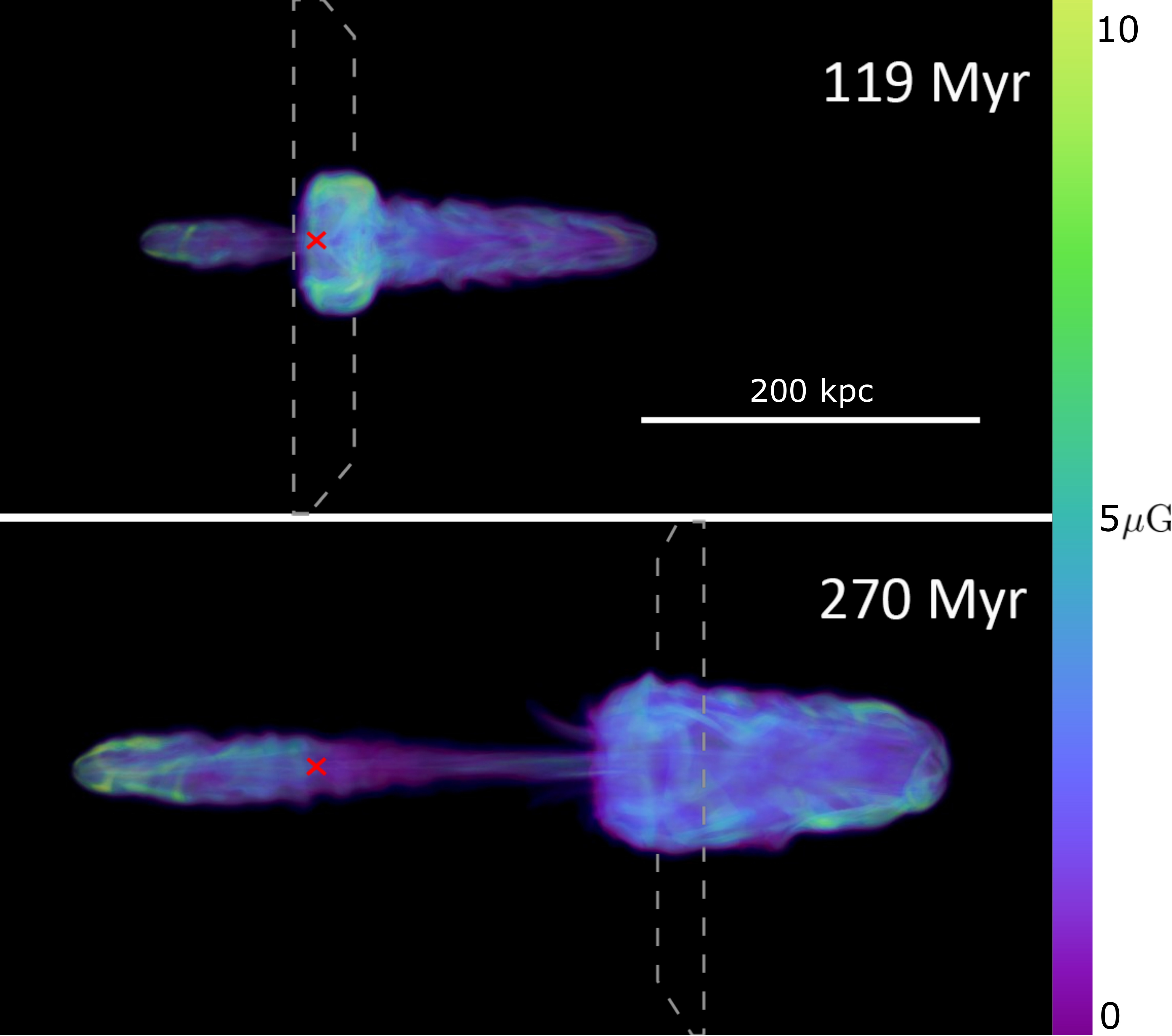}
\caption{Volume rendering of the magnetic field intensity in the {\bf J3S2} simulation at the times in Figure \ref{fig:m2evolve}. AGN is marked with a red ``x.'' As in previous figures, the location of the external ICM shock is outlined in dashed gray.}
\label{fig:m2Bmag}
\end{figure}
%%%%%%%%%%%%%%%%%%%%%%%%%%%%
%%%%%%%%%%%%%%%%%%%%%%%%%%%%

The dynamical distinctions between {\bf J3S2} and {\bf J3S4a} are even more obvious on the downwind side, and especially at later times. For instance, in Figure \ref{fig:m2evolve} it is apparent that the $M_{si} = 2$ external (ICM) shock at $t = 270$ Myr in {\bf J3S2} was $\sim 200$ kpc behind the downwind jet head. In contrast, at the earlier $t = 230$ Myr the $M_{si} = 4$ external shock in Figure \ref{fig:m4evolve} was overtaking the full downwind RG structure, and the vortex ring formation had disrupted the downwind jet propagation. In fact, the external ICM shock in {\bf J3S2} never did during the simulation overtake the end of the downwind jet. Figure \ref{fig:m2evolve} reveals a vortex ring surrounded the downwind RG lobe, and a high pressure, high density region surrounded the downwind jet at $t = 270$ Myr. There was also substantial ICM plasma inside the downwind lobe, carried forward by the shock internal to the cavity. However, it remains clear that neither the jet nor the downwind lobe had been disrupted. As a consequence, and again in contrast to {\bf J3S4a}, the vortex ring never developed into an isolated structure.

Additional similarities and differences between {\bf J3S2} and {\bf J3S4a} can be seen by comparing the magnetic field volume renderings in Figures \ref{fig:m2Bmag} and Figure \ref{fig:m4Bfield}. In both cases peak magnetic field strengths several times greater than those in the jets resulted from stretching by vortical motions. Allowing for differences in jet behaviors, the magnetic field distribution at $t = 119$ Myr in {\bf J3S2} was qualitatively similar to that at $t = 92$ Myr in {\bf J3S4a}. In particular, both cases exhibit substantial magnetic field amplification in their nascent vortex ring structures, along with more modest field enhancements in the RG lobes. But, again at the later times ($t = 270$ Myr in Figure \ref{fig:m2Bmag} and $t = 230$ Myr in Figure \ref{fig:m4Bfield}) the differences from the relative rate of shock propagation and the strength of the vortex ring in the two simulations are obvious. Both RG lobes in Figure \ref{fig:m2Bmag} exhibited substantial magnetic field amplification by this time. In contrast, and as described above, the upwind side of {\bf J3S4a} was completely absent, and the downwind jet in {\bf J3S4a} had been disrupted well before this time by formation of the vortex ring. The {\bf J3S4a} vortex ring was by then isolated from the AGN and being advected away by the post-shock wind. One of the most obvious regions of amplified magnetic field in the downwind region of {\bf J3S4a} at late times was, in fact, the vortex ring. 

\begin{figure*}
\includegraphics[width=17.5cm]{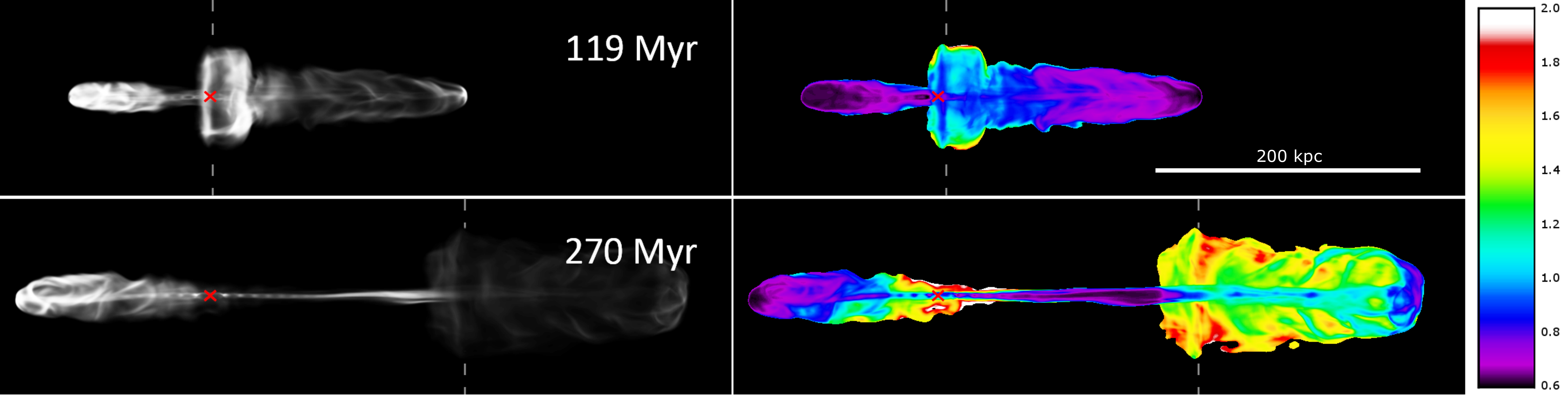}
\caption{ Synchrotron images from {\bf J3S2} at the times in Figure \ref{fig:m2evolve}. Resolution is 0.5 kpc. The AGN jet axis and shock normal are in the plane of the sky. Left: 150 MHz intensity, with the brightest pixel in each image approximately the highest intensity at a given time excluding the jet launch cylinder. Right: 150/600 MHz spectral index for regions above 0.125\% of the peak intensity at 150 MHz. Spectral index scale is on the far right. At launch the jet spectral index was $\alpha = 0.6$. AGN is marked with a red ``x.'' The location of the external ICM shock is denoted by a dashed gray line.}
\label{fig:m2-synch}
\end{figure*}

Figure \ref{fig:m2-synch} illustrates the synchrotron surface brightness images of {\bf J3S2} at 150 MHz along with the associated $\alpha_{150/600}$ spectral distributions at $t = 119$ Myr and $t = 270$ Myr. It is remarkable, given the evidence in Figure \ref{fig:m2Bmag} for substantial magnetic field amplification in the downwind (right side) RG lobe, that the 150 MHz surface brightness from that structure was mostly relatively low at both times shown. In fact at $t = 270$ Myr the downwind jet exhibited significantly higher surface brightness than the lobe on that side, despite the relatively weaker magnetic field in the jet. Indeed, although some volumes in {\bf J3S2} with strongly amplified magnetic fields were radio bright, such as the nascent vortex ring at $t = 119$ Myr and the upwind lobe at $t = 270$ Myr, the 150 MHz surface brightness was not a consistent indicator of magnetic field strength patterns. This is in contrast to what we found for {\bf J3S4a}. 
%%%%%%%%%%%%%%%%%%%%%%%%%%
%%%%%%%%%%%%%%%%%%%%%%%%%%
%%%%%%%%%%%%%%%%%%%%%%%%%%
%%%%%%%%%%%%%%%%%%%%%%%%%%

Several features revealed in Figure \ref{fig:m2evolve} and the right side of Figure \ref{fig:m2-synch} combine to explain this different result. 
Looking first at Figure \ref{fig:m2evolve}, we see that at both times, but especially at the later time, the jet mass fraction distribution in the downwind lobe of {\bf J3S2} was generally lower than in the upwind lobe. This came from the fact that the post-shock wind upwind of the vortex ring had blown away most of the upwind AGN plasma that was deposited at early times, leaving only relatively fresh, unmixed AGN plasma. However, all of the downwind AGN plasma had been retained and had become entrained with substantial ICM plasma in response to the penetration of the ICM driven shock. The jets themselves, on both sides, actually remained essentially fresh, unmixed AGN injected plasma. 

This explanation is reinforced by the $\alpha_{150/600}$ distributions on the right in Figure \ref{fig:m2-synch}. Although at $t = 119$ Myr both RG lobes exhibit spectra similar to the injected, $\alpha = 0.6$, virtually all of the high surface brightness regions in the upwind lobe are flatter (so dominated by younger CRe) than any portion of the downwind lobe except its very tip, where the downwind jet was actively depositing fresh AGN CRe within $\sim 10$ Myr of its injection at the AGN . The spectrum of the nascent vortex ring is notably steeper, with $\alpha_{150/600} \ga 1$, but, since this region also contained the strongest magnetic fields at this time it was radio bright. At the later time, $t = 270$ Myr, $\alpha_{150/600} \ga 1.2$ everywhere on the downwind side except for the jet itself and for small regions of freshly deposited CRe in the head of the lobe. This reflects the fact that the emitting CRe population in the lobe on this side had accumulated for almost 300 Myr, so was significantly aged. Recall that the radiative cooling times of the radio visible CRe, $\tau_{rad} \la 100$ Myr. Since the magnetic fields in the downwind lobe were not generally stronger than those in the upwind lobe, the aged, steeper-spectrum downwind CRe population led to reduced synchrotron surface brightness along with steeper synchrotron spectra.
%%%%%%%%%%%%%%%%%%%%%%%%%%
%%%%%%%%%%%%%%%%%%%%%%%%%%
\begin{figure*}
\centering
\includegraphics[width=9cm]{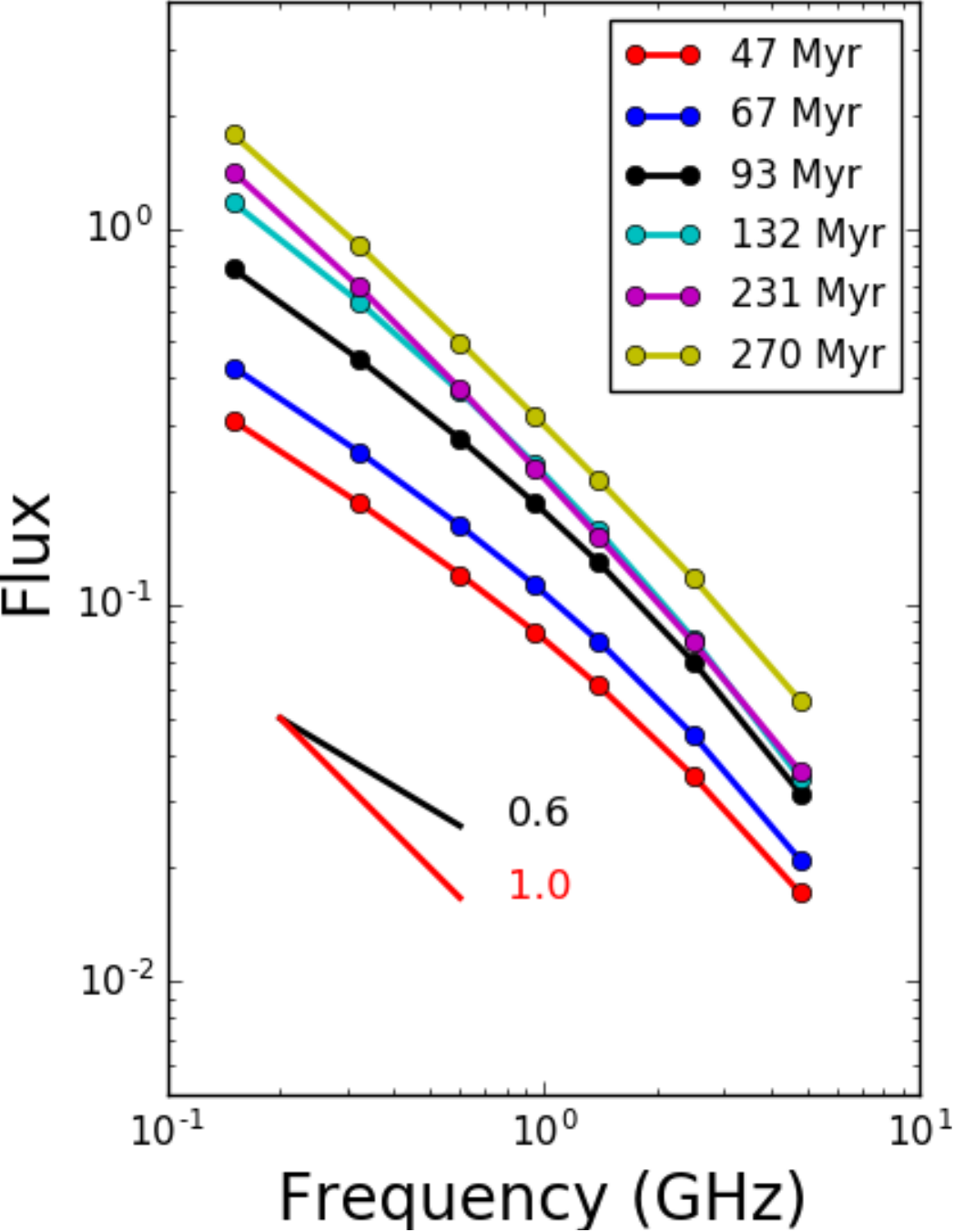}
\caption{Evolution of the {\bf J3S2} integrated synchrotron spectrum.}
\label{fig:m2-spec}
\end{figure*}
%%%%%%%%%%%%%%%%%%%%%%%%%%
%%%%%%%%%%%%%%%%%%%%%%%%%% 

The properties just outlined come together to account for the source-integrated spectrum of {\bf J3S2} illustrated at multiple times in Figure \ref{fig:m2-spec}. Before $t \sim 100$ Myr the integrated fluxes and spectra are rather similar to those in {\bf J3S4a} and show only modest evidence for aging, reflecting the minor differences in the RG evolutions to that point and steady contributions of fresh CRe to dominant emission regions. At later times, although the integrated spectra of both objects steepen and are similar at low frequencies, the spectra of {\bf J3S4a} is significantly steeper at high frequencies, reflecting the importance to {\bf J3S4a} of the isolated vortex ring with its aging CRe population. 

\subsection{Simulation J3S22: $\theta_j =15^{\degree}$, $M_{ji} = 3.5$, $M_{si} = 2.25$}
\label{subsec:misalign}
Both of the {\bf J3S4a} and {\bf J3S2} simulations included exact alignment between the AGN jet axis and the ICM shock normal ($\theta_j = 0$). The simulation {\bf J3S22}, with $\theta_j = 15^{\degree}$, breaks the axisymmetry of those aligned interactions, but still develops behaviors that are similar in many ways to the aligned interactions. Here we point out some obvious differences due to the symmetry breaking, but defer to work in preparation, \cite{ONeill19b}, analysis of the more general problem of jets interacting with cross winds at arbitrary angle, $\theta_j$ to make twin tailed structures. 

Except for alignment, the jet properties in {\bf J3S22} were exactly the same as those in {\bf J3S4a} and {\bf J3S2}. The incident shock was given a Mach number, $M_{si} = 2.25$, which, from equations \ref{eq:AdvanceRateModified} and \ref{eq:stopjet} for aligned jet--shock encounters we estimated, after adjusting for the alignment, should just stop forward progress of the upwind jet (see also Figure \ref{fig:vhead_test}). In that regard {\bf J3S4a} is a closer analogy to {\bf J3S22} than {\bf J3S2}, in which the upwind jet was not reversed. On the other hand, the high pressure in the post-shock wind in {\bf J3S4a} led the jet flows to become subsonic, while equation \ref{eq:subsonicjet} suggested that the jets in {\bf J3S22} would remain supersonic. That expectation was confirmed by the simulations.

Otherwise, we anticipated two principal differences between {\bf J3S4a} and {\bf J3S22}, both coming from the symmetry break. First, we expected (and confirmed) the misalignment of the upwind jet in {\bf J3S22} would cause it to be sharply deflected into the downstream direction rather than directly reversed, thus making two, close, downwind-facing jets/tails. Second, we expected (and confirmed) the high pressure (and high density) structure developing just upwind of the shock-induced vortex ring would disrupt the downwind jet more by strong deflection than by strong pinching. 
Beyond the distinctions outlined here, there were no characteristic dynamical patterns in {\bf J3S22} that were not qualitatively similar to those found in {\bf J3S4a}. Consequently, we do not present a detailed analysis of the former. 
%%%%%%%%%%%%%%%%%%%%%%%%%%%%
%%%%%%%%%%%%%%%%%%%%%%%%%%%%
\begin{figure*}
\centering
\includegraphics[width=14cm]{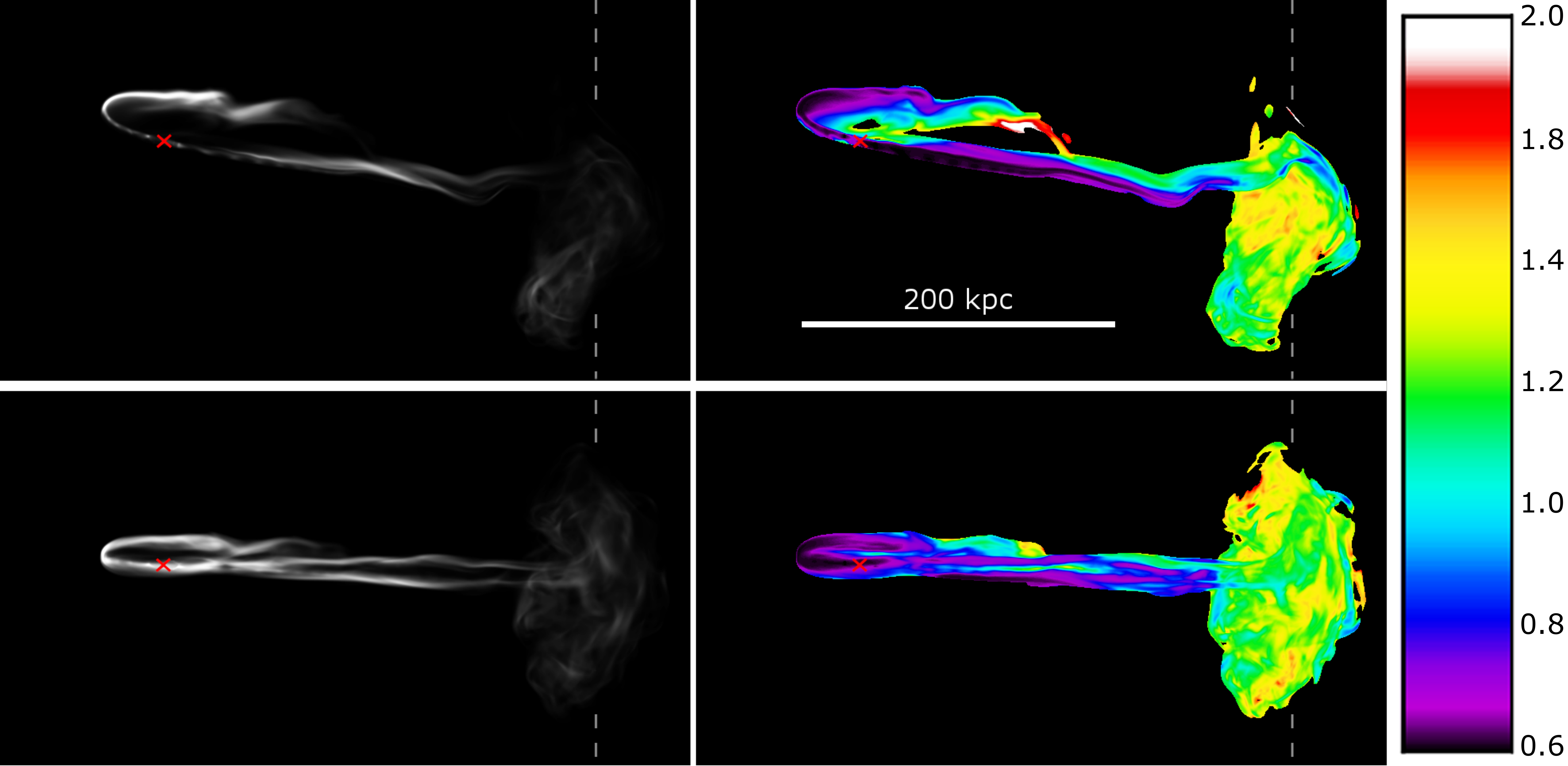}

\caption{ 150 MHz images (left) and 150 MHz/600 MHz spectral index distributions (right) from {\bf J3S22} at $t = 295$ Myr. The shock normal is in the plane of the sky. Top: AGN jet axis is in the plane of the sky. Bottom: The jet--shock normal plane is rotated around the shock normal by $75^{\degree}$ from the sky. Spectral index color bar is on the far right. The AGN position is marked by a red ``x.'' The location of the external ICM shock is denoted by a dashed gray line.}
\label{fig:misalign}
\end{figure*}
%%%%%%%%%%%%%%%%%%%%%%%%%%%%
%%%%%%%%%%%%%%%%%%%%%%%%%%%%

However, in order to provide some concrete illustration of the consequences of symmetry breaking, we present 150 MHz synchrotron images and associated spectral index maps of {\bf J3S22} at $t = 295$ Myr in Figure \ref{fig:misalign}. Two viewing orientations are shown, both with the shock normal in the plane of the sky. In the upper images the (oblique) jet axis is also in the plane of the sky. This reveals the full physical separation of the two tails, which is generally $\la 30$ kpc even several hundred kpc downwind from the AGN. In the images, the upwind (leftward from the `x') jet is aimed upwards from horizontal by $15^{\degree}$, while the downwind jet is aimed $15^{\degree}$ downwards from horizontal. The lower panel differs only in that the plane containing the jet axis has been rotated by $75^{\degree}$ around the shock normal (still in the plane of the sky), so that the two tails now appear more nearly overlapping. By this time the shock itself is well to the right of all the synchrotron emitting plasma and out of the image field of view. The vortex ring is well developed, although weaker and, because of less magnetic field amplification, not as radio luminous as it was in {\bf J3S4a}. As in {\bf J3S4a} the downwind jet has been disrupted near the vortex ring. Although the condition derived in \S \ref{subsec:M4} was only marginally satisfied, the pressure imbalance across the jet as it interacted with the vortex flow was still sufficient to deflect the jet substantially, leading to its disruption. One consequence of the break in symmetry can be seen as an apparent bifurcation in the synchrotron emissions from the jet, seen most obviously in the bottom left panel of figure \ref{fig:misalign}. Ram pressure across the jet exerted by the wind creates shear along the jet boundary that causes the initially torroidal field to become predominantly poloidal. Due to the jet bending, the cylindrical symmetry of the jets is broken and this poloidal magnetic field forms two distinct filaments on opposite sides of the jet. Since the synchrotron emissivity is especially sensitive to magnetic field strength, the resultant image emphasizes this pattern. More detailed analysis of this effect is left to \cite{ONeill19a}.

Qualitatively the observable structures in {\bf J3S22} at $t = 295$ Myr are rather similar to those at $t = 230$ Myr in {\bf J3S4a}, once we account for differences in wind velocities and their influence on the displacement of the vortex ring over time. The synchrotron spectrum in the vortex ring here is steeper than it was in {\bf J3S4a}, because the weaker magnetic field also means that emissions at a given frequency come from higher energy CRe. Thus, the spectra are more sensitive to radiative cooling, which again is in this simulation dominated by inverse Compton emissions.

We conclude that for moderately misaligned jet shock interactions the simple, aligned case analysis provides a reasonable dynamical template, although there are some potentially observable consequences of the misalignment. Most significant, of course, the misaligned interaction that reverses the upwind jet really leads to the formation of close tails, rather than a coaxial downwind jet structure.
Whether an observer would see the example source in Figure \ref{fig:misalign} as a single, one sided tail or twin tails would clearly depend on the projection of the source in the sky, on the actual distance to the source and on the effective spatial resolution available to the observer. At z = 0.2 the maximum physical separation ($\sim 30$ kpc) corresponds roughly to 7.5 arc seconds. For most projections the apparent separation would be significantly less than this, of course. Observational separation would likely be challenging, but possible \citep[see, e.g.][]{deGregory17}.

\section{Summary}
\label{sec:Summary}

We have reported a study of the interactions between lobed radio galaxies (RG) formed by active supersonic, light jets and plane ICM shocks when the radio galaxy jet axis and the shock normal are virtually aligned. Our study utilizes 3D MHD simulations designed to test and extend insights coming from simple analytic analyses. The dynamics of those interactions can be decomposed into the impulsive pressure, density and velocity changes introduced to the RG by the shock discontinuity, followed by the subsequent longer term interactions between the RG plasma and the RG jets with the post-shock wind. 

\subsection*{Consequences of Shock passage}

As previous studies have emphasized, an ICM shock impacting a low density RG lobe propagates very rapidly through the lobe, drawing in external ICM plasma and generating strong shear along the lobe perimeter. The most distinctive feature of this shock penetration is the development through boundary shear of a toroidal vortex ring that is carried downwind in the post-shock flow. This vortex can significantly amplify magnetic fields from the RG lobe and, if strong enough, disrupt the RG jets downwind by way of a high pressure feature that forms along the vortex's toroidal axis.

Our simulations also included a population of relativistic electrons that we used to explore radio synchrotron emissions during the shock--RG encounters. We found that magnetic field amplification coming especially from field line stretching where vortical motions were strong could substantially enhance radio emissions in structures produced during the shock encounters. This consequence appears to be especially significant in the large vortex ring structures formed in post-shock flows.

\subsection*{Consequences of Interaction with a wind}

The predominant consequence of prolonged post-shock wind interaction is the influence of ram pressure on the trajectory of the jets. When, as in this study, the post-shock wind and the jets are aligned, the most direct influence can be characterized in terms of a reduction in the rate at which the upwind jet penetrates into the ICM. For a given set of jet and wind parameters the resulting behaviors do not depend essentially on the source of the wind, but only on its properties with respect to the RG and its host galaxy. The shock applied in this study is only one means to generate the appropriate wind.

A sufficiently strong headwind can reverse the propagation of the upwind jet (or, if they are misaligned, sharply bend that jet into a tail close to the downwind jet/tail). The details of the modifications to jet propagation are sensitive to such things as the ratio of the internal Mach number of the jet and the Mach number of the ICM shock. Thus, the outcomes offer potential metrics of the RG and the ICM.

All these outcomes are rather distinctive and suggest that detailed observations of deformed RGs in clusters could provide valuable insights into both the physics of RG-shock interactions and also the character of large scale ICM dynamical flow patterns.

\acknowledgements

This work was supported at the University of Minnesota by NSF grant AST1714205 and by the Minnesota Supercomputing Institute. CN was supported by an NSF Graduate Fellowship under Grant 00003920 as well as with a travel grant through the School of Physics and Astronomy at the University of Minnesota. We thank numerous colleagues, but especially Larry Rudnick and Avery F. Garon for encouragement and feedback.

\bibliography{ShockedRGs-Aligned}

\end{document}